\documentclass[twoside]{article}
\usepackage{color}
\usepackage{amssymb,amsmath,amsthm} 
\usepackage{verbatim}
\usepackage{shadow}

\usepackage{boxedminipage}
\usepackage{fancybox}
\usepackage{graphicx,float}
\usepackage{newcent}
\usepackage{amssymb}%
\usepackage[hidelinks]{hyperref}
\usepackage{fancyhdr}
\usepackage{xcolor}
\usepackage{amsfonts}
\usepackage{graphicx,epstopdf}
\usepackage[caption=false]{subfig}
\usepackage[official,right]{eurosym}
\usepackage{booktabs}

\usepackage{color}
\usepackage{amssymb,amsmath,amsthm} 
\usepackage{verbatim}
\usepackage{shadow}

\usepackage{boxedminipage}
\usepackage{fancybox}
\usepackage{graphicx,float}
\usepackage{newcent}
\usepackage{amssymb}%
\usepackage{fancyhdr}
\usepackage{xcolor}
\usepackage{booktabs}
\numberwithin{equation}{section}

\newtheorem{theorem}{Theorem}[section]
\newtheorem{lem}{Lemma}[section]
\newtheorem{pro}{Proposition}[section]
\newtheorem{cor}{Corollary}[section]
\newtheorem{rem}{Remark}[section]
\newtheorem{rems}{Remarks}[section]
\newtheorem{ex}{Example}[section]
\newtheorem{defi}{Definition}[section]
\newtheorem{hyp}{Assumption}[section]

\newcommand{\bt}{\begin{theorem}}
\newcommand{\et}{\end{theorem}}
\newcommand{\bl}{\begin{lem}}
\newcommand{\el}{\end{lem}}
\newcommand{\bp}{\begin{pro}}
\newcommand{\ep}{\end{pro}}
\newcommand{\bcor}{\begin{cor}}
\newcommand{\ecor}{\end{cor}}

\newcommand{\bd}{\begin{defi} \rm }
\newcommand{\ed}{\end{defi}}
\newcommand{\brem }{\begin{rem} \rm }
\newcommand{\erem }{\end{rem}}
\newcommand{\brems }{\begin{rems} \rm }
\newcommand{\erems }{\end{rems}}
\newcommand{\bhyp }{\begin{hyp} \rm }
\newcommand{\ehyp }{\end{hyp}}
\newcommand{\bex}{\begin{ex} \rm }
\newcommand{\eex}{\end{ex}}

\newcommand{\btT}{\sigma_{\Td}(t)}

\newcommand{\btU}{\sigma_{\Ud}(t)}

\newcommand{\hbtT}{\wh{\sigma}_{\Td}(t)}

\newcommand{\hbtU}{\wh{\sigma}_{\Ud}(t)}

\newcommand{\Xbeta}{X^{\beta}}

\newcommand{\mart}{\simeq}

\newcommand{\cq}{q}

\newcommand{\diff}{\mathop{}\!d}
\newcommand{\wh}{\widehat}
\newcommand{\wt}{\widetilde}

\newcommand{\lqt}{\lambda^q_t}

\newcommand{\rhodf}{\rho_{12}}
\newcommand{\rhod}{\rho_{13}}
\newcommand{\rhof}{\rho_{23}}

\newcommand{\rd}{r^d}
\newcommand{\rf}{r^f}

\newcommand{\xd}{\bar{r}^d}
\newcommand{\xf}{\bar{r}^f}

\newcommand{\phiv}{\varphi}
\newcommand{\phid}{\varphi^d}
\newcommand{\phif}{\varphi^f}

\newcommand{\Bd}{B^d}

\newcommand{\Bf}{B^f}
\newcommand{\Rd}{R^d}
\newcommand{\Rf}{R^f}

\newcommand{\Td}{T}

\newcommand{\Ud}{U}

\newcommand{\barsig}{\bar{\sigma}}

\newcommand{\Zone}{Z^1}
\newcommand{\Ztwo}{Z^2}
\newcommand{\Zthree}{Z^3}

\newcommand{\whZone}{\wh{Z}^1}
\newcommand{\whZtwo}{\wh{Z}^2}
\newcommand{\whZthree}{\wh{Z}^3}

\newcommand{\Fd}{F^d}

\newcommand{\Ff}{F^f}

\newcommand{\Ffq}{F^{f,\cq}}

\newcommand{\nud}{\nu^d}

\newcommand{\nuf}{\nu^f}

\newcommand{\nufq}{\nu^{f,\cq}}

\newcommand{\ff}{{\mathbb F}}
\newcommand{\pp}{{\mathbb P}}

\newcommand{\E}{{\mathbb E}}

\newcommand{\Q}{\mathbb{Q}}

\newcommand{\EQ}{{\E}_{\Q}}

\newcommand{\whQ}{\wh{\Q}}
\newcommand{\EwhQ}{{\E}_{\whQ}}

\newcommand{\wtQ}{\wt{\Q}}

\newcommand{\Noti}{P}
\newcommand{\DS}{{\rm \bf DS}}

\newcommand{\Swapb}{\DS^{\beta,\gamma,\kappa}}

\newcommand{\SwapT}{\DS^{\gamma,\kappa}}

\newcommand{\cE}{\mathcal E}
\newcommand{\cF}{\mathcal F}

\definecolor{mybl}{rgb}{0,0.29,0.69}

\DeclareRobustCommand{\qvar}[2]{\ensuremath{\langle{#1} \rangle}_{#2}}

\ifpdf
  \DeclareGraphicsExtensions{.eps,.pdf,.png,.jpg}
\else
  \DeclareGraphicsExtensions{.eps}
\fi

\makeatletter
\def\mathcolor#1#{\@mathcolor{#1}}
\def\@mathcolor#1#2#3{%
  \protect\leavevmode
  \begingroup
    \color#1{#2}#3%
  \endgroup
}
\makeatother

\title{{\Large \bf Choice of Collateral Currency in Differential Swaps}\vskip 20 pt }

\author{Yining Ding$\,^{a}$, Ruyi Liu$\,^{b}$ and Marek Rutkowski$\,^{a,c}$ \\ \\ \\ \\
\\
$^{a\,}$School of Mathematics and Statistics, University of Sydney \\
Sydney, NSW 2006, Australia \\ \\
$^{b\,}$School of Mathematics and Statistics, University of New South Wales \\
Sydney, NSW 2033, Australia \\ \\
$^{c\,}$Faculty of Mathematics and Information Science, Warsaw University of Technology \\
00-661 Warszawa, Poland
}

\date{\vskip 25 pt October 4, 2026 \vskip 20 pt}

\begin{document}

\maketitle

\begin{abstract}
The role of collateral in derivative pricing has evolved beyond credit risk mitigation, particularly following the global financial crisis, when funding costs and basis spreads became central to
valuation practices. This development coincided with the transition from the London Interbank
Offered Rate (LIBOR) to risk-free rates (RFRs) and the increasing standardization of collateralised
trading. We study the valuation and hedging of a class of differential swaps referencing backward-looking averages of overnight rates, with Secured Overnight Financing Rate (SOFR) swaps appearing as a particular instance. The focus is on the impact of the collateral currency. Extending the earlier results of Bickersteth et al. [Math. Finance 36 (2026), pp.~180--202], we allow the collateral account to be denominated in a currency different from that of the contractual cash flows and derive explicit pricing and hedging strategies using a futures-based replication approach.

We show that the choice of collateral currency can have a non-trivial effect on both valuation and risk management through discounting adjustments and payment-date weights. Numerical experiments validate the closed-form formulas, demonstrate that collateral effects can produce valuation adjustments, and confirm the effectiveness of the futures hedges under discrete rebalancing.
\end{abstract}

\vskip 40 pt
\noindent \textbf{Keywords:}
collateral currency, differential swaps, SOFR, \euro STR, cross-currency valuation, interest-rate futures, futures-based hedging \\
\noindent \textbf{MSC:}
60H10, 60H30, 91G30, 91G40


\newpage

\section{Introduction}

The global financial crisis of 2007--2009 and the subsequent move toward central clearing and strengthened margining practices contributed to the market-wide dominance of collateralised trading in over-the-counter derivatives. In parallel, international standards on margin requirements for non-centrally cleared derivatives emphasise variation margin as a key tool for limiting counterparty exposure and systemic risk \cite{bcbsiosco2015margin}. Beyond its role in credit risk mitigation, collateralisation has economically meaningful pricing implications: collateral provisions can affect swap rates and, more broadly, the valuation of interest-rate derivatives \cite{johannesSundaresan2007collateralSwapRates}. Fixed-income markets consequently adopted a post-crisis architecture in which discounting is aligned with collateral remuneration and curve construction is explicitly multi-curve \cite{bianchetti2009twoCurves,henrard2010ironyCrisis,mercurio2010modernLMM}.

This paper studies a question that naturally arises in this environment: \emph{how does the choice of collateral currency affect the valuation and hedging of interest-rate derivatives?} A Credit Support Annex (CSA) may permit collateral to be posted in a currency different from the settlement currency of the target trade. For example, a USD SOFR trade may be collateralised with EUR cash remunerated at the euro short-term rate (\(\text{\euro STR}\)) \cite{laurentAmzelekBonnaud2014overview}. Here USD and EUR are used for illustration, the currency pair may be replaced by any other currency--RFR pair.

These developments also prompted a broader literature on collateralised valuation. Collateral remuneration determines discounting under idealised assumptions, while other frameworks incorporate imperfect collateralisation, haircuts, asymmetries, and funding considerations \cite{brigoMoriniPallavicini2013ccrcf,fujii2013asymmetricImperfectCollateral,laurentAmzelekBonnaud2014overview}. There is also an active debate on the interpretation and accounting status of funding valuation adjustments \cite{andersenDuffieSong2019fva,hullWhite2014fvaFairValue}.

The distinction between the currency of contractual cash flows and the currency of collateral is central to multi-currency collateralised valuation. Early contributions developed multiple-curve and cross-currency frameworks with collateral and dynamic basis spreads \cite{fujii2010note,fujii2010multi,FujiiShimadaTakahashi2011Dynamic,FujiiTakahashi2016Benchmark}; subsequent work treated imperfect collateralisation, arbitrary numeraires, and general cross-currency valuation and hedging \cite{fujii2013asymmetricImperfectCollateral,JK2020,Gnoatto2021multi,Gnoatto2023cchjm}. When a CSA permits substitution among several eligible collateral currencies, the posting party may also possess a cheapest-to-deliver option \cite{Fujii2011Choice,Wolf2022}. The present paper fixes the selected collateral currency over the life of the trade and therefore gives a static comparison of currency-specific collateral carry.

The market context is further shaped by benchmark reform and the transition from IBORs to overnight risk-free rates \cite{duffieStein2015benchmarks}. The backward-looking nature of compounded overnight rates has led to dedicated pricing and term-structure models, including affine specifications for backward-looking rates and overnight-rate term structures \cite{Lyashenko2019SOFR,fontana2023affine,FGS2023}. Complementary studies use SOFR futures to construct the term structure and to capture scheduled discontinuities and stochastic-volatility effects \cite{Skov2021rer,SchloglSkovSkovmand2024,Brace2024}. Within this RFR-based market structure, exchange-traded futures linked to SOFR and \(\text{\euro STR}\) provide natural hedging instruments. Recent work develops futures-based pricing and hedging methods for SOFR derivatives and cross-currency products referencing backward-looking rates \cite{Bickersteth2026SOFR,Ding2024wp}. Since futures are settled daily, classical forward--futures distinctions and convexity effects remain relevant both for valuation and for the construction of self-financing hedges \cite{coxIngersollRoss1981forwardFuturesRelation,jarrowOldfield1981forwardsFutures}.

Our main object is a class of USD-settled differential swaps referencing backward-looking SOFR and \(\text{\euro STR}\) averages. The constant \(\gamma\) is a contractual scaling parameter for the foreign floating-rate component: \(\gamma=0\) gives a standard SOFR swap, \(\gamma=1\) gives unit exposure to the \(\text{\euro STR}\) average, and any other fixed value rescales that exposure. A non-zero \(\gamma\) may in particular represent a predetermined exchange rate, in which case the contract is a quanto-style differential swap. The target trade may be collateralised in either USD or EUR, with an exogenous proportional collateralisation assumption.

Within this setup, our contributions are as follows. First, we formulate futures-based self-financing strategies under explicit collateral funding and settlement conventions. We distinguish the overnight benchmark accounts from the tradable money-market accounts and the actual account balances from the hedger's net wealth. This identifies the funding of both posted and received collateral and yields the discounting rule under USD and EUR collateral. Second, under a joint Vasicek--Garman--Kohlhagen specification, we derive closed-form prices for single- and multi-period differential swaps. The formulas separate collateral-specific discounting weights from the contractual quanto correction and are expressed using traded SOFR and \(\text{\euro STR}\) futures rates. Third, we derive explicit replicating strategies in these futures. Finally, we quantify the two channels separately. The numerical study compares USD and EUR collateral for the same legal contract, measures the contractual quanto adjustment, and evaluates the futures hedges under discrete rebalancing.

The Gaussian specification is deliberately parsimonious. The unshifted Vasicek models cannot fit arbitrary observed initial term structures, although deterministic shifts may be introduced to fit the initial curves without altering the one-factor diffusion structure. One-factor short rates do not capture slope and twist risks. The theoretical role of the closed-form formulas is to isolate the collateral-carry and contractual cross-currency channels and to provide exact benchmarks for numerical implementations of richer models. The self-financing framework and the pricing decomposition are not tied to the Vasicek specification; in broader affine or multi-factor Gaussian models, the same structure is retained whenever the required conditional exponential moments are available.

The remainder of the paper is organised as follows. Section~\ref{sec:collat-funding} introduces the model-free collateral and futures-settlement conventions, the self-financing wealth dynamics, and the pricing martingale measure. Section~\ref{sec:cc-tsm} specifies the cross-currency Gaussian model and derives the dynamics of the traded futures. Section~\ref{sec:pricing} gives the single- and multi-period pricing results under USD and EUR collateral. Section~\ref{sec:hedging} constructs the corresponding futures-based replicating strategies. Section~\ref{sec:numerics} validates the closed-form results and separates the collateral-carry effect, the contractual quanto adjustment, and the discrete implementation error of the hedges.

\section{Collateralised Trading Strategies and Pricing Measure}\label{sec:collat-funding}

Throughout the paper, the domestic currency is USD, the foreign currency is EUR, and \(Q_t\) denotes the number of units of domestic currency per unit of foreign currency. Although USD and EUR are used for concreteness, the currency pair and their respective overnight benchmarks may be replaced by any other currency--RFR pair.

This section fixes notation and gives a model-free setup for collateralisation, funding, and futures-based self-financing trading in this cross-currency market. We distinguish the overnight benchmark rates from the tradable money-market accounts, specify how foreign futures variation margin is converted into domestic currency, and derive a unified effective rate \(r^{\beta,i}\) when collateral is denominated in either USD or EUR, where \(i\in\{d,f\}\) identifies the collateral currency.

We first summarise the relevant market conventions and linked instruments, then formalise admissible futures strategies for collateralised contracts together with their self-financing condition, and finally introduce the pricing martingale measure associated with the effective rate \(r^{\beta,i}\).

The setup of this section is model-free and does not rely on a specific term-structure specification. Pricing is formulated under a pricing martingale measure $\wtQ$ for the futures gains processes (see Definition~\ref{def:pmm}). The collateral specification determines the effective discounting account. In Section~\ref{sec:cc-tsm}, we specify a cross-currency term-structure model under the domestic martingale measure $\Q$ and verify that $\Q$ is indeed a pricing martingale measure for the collateralised futures strategies, see Proposition~\ref{pro3.1}.

All random objects are defined on a filtered probability space \((\Omega,\cF,\mathbb{F},\pp)\) where \(\mathbb{F}=(\cF_t)_{t\ge0}\) satisfies the usual conditions. For each contract, all trading and settlement dates are contained in a fixed finite horizon, and all processes introduced below are assumed to be \(\mathbb{F}\)-adapted.

\subsection{Overnight benchmarks, money-market accounts, and related futures}  \label{sec:conventions}

Throughout this section, we fix an accrual period $[\Ud,\Td]$ with $0\le \Ud<\Td$ and set
$\delta:=\Td-\Ud$. For notational convenience, whenever no confusion arises, we suppress the dependence of futures quotes on $[\Ud,\Td]$. The domestic overnight benchmark is
SOFR and the foreign overnight benchmark is \(\text{\euro STR}\) with respective
compound averages given by the following definition, which is based on a continuous time
proxy for the actual market conventions.

\bd\label{def:benchmarks}
Let the $\mathbb{F}$-adapted processes $\rd$ and $\rf$ represent the instantaneous SOFR and
\(\text{\euro STR}\) rates, respectively. The continuously compounded \emph{SOFR account} $\Bd$ satisfies
\[
\Bd_t=\exp\Big(\int_0^t \rd_u\diff u\Big),\qquad t\ge 0,
\]
and the backward-looking \emph{SOFR Average} over $[\Ud,\Td]$ is
\[
\Rd(\Ud,\Td):=\frac{1}{\delta}\Big(\exp\Big(\int_{\Ud}^{\Td}\rd_u\diff u\Big)-1\Big)
=\frac{1}{\delta}\Big(\frac{\Bd_{\Td}}{\Bd_{\Ud}}-1\Big).
\]
Similarly, the continuously compounded \emph{\(\text{\euro STR}\) account} $\Bf$ satisfies
\[
\Bf_t=\exp\Big(\int_0^t \rf_u\diff u\Big),\qquad t\ge 0,
\]
and the backward-looking \emph{\(\text{\euro STR}\) Average} over $[\Ud,\Td]$ is
\[
\Rf(\Ud,\Td):=\frac{1}{\delta}\Big(\exp\Big(\int_{\Ud}^{\Td}\rf_u\diff u\Big)-1\Big)
=\frac{1}{\delta}\Big(\frac{\Bf_{\Td}}{\Bf_{\Ud}}-1\Big).
\]
\ed

Both $\Rd(\Ud,\Td)$ and $\Rf(\Ud,\Td)$ are backward-looking so, in particular, they are $\cF_{\Td}$-measurable but not
$\cF_t$-measurable for $t<\Td$. Futures written on these realised averages will be our primary hedging instruments.

The processes $r^d$ and $r^f$ are overnight benchmark rates used to define the reference accruals. They need not coincide with the borrowing and investment rates available to a bank or corporate hedger, which generally include currency-specific funding and liquidity spreads. The accounts $B^d$ and $B^f$ in Definition~\ref{def:benchmarks} are therefore benchmark accumulation processes and are not postulated as additional funding instruments. To represent attainable cash funding and investment, we separately introduce the tradable domestic and foreign money-market accounts $\bar B^d$ and $\bar B^f$ by
\begin{equation}\label{rty1}
\diff \bar B^d_t=\xd_t\bar B^d_t\diff t,
\qquad
\diff \bar B^f_t=\xf_t\bar B^f_t\diff t,
\qquad \bar B^d_0=\bar B^f_0=1,
\end{equation}
where
\[
\xd_t:=\rd_t+\alpha^d_t,
\qquad
\xf_t:=\rf_t+\alpha^f_t.
\]
The spreads $\alpha^d$ and $\alpha^f$ are deterministic and integrable on every finite time interval.

\bhyp\label{ass:direct-funding}
The hedger can borrow and invest directly in both currencies through the money-market accounts $\bar B^d$ and $\bar B^f$. Borrowing and investment in currency $i \in \{d,f\}$ accrue at the same rate $\bar r^i$. All USD cash balances are funded or invested through $\bar B^d$, so the domestic hedge funding account and rate satisfy
\[
B^h=\bar B^d,\qquad r^h=\xd.
\]
\ehyp

Assumption~\ref{ass:direct-funding} specifies the cash funding facilities available to the hedger. In particular, EUR collateral can be posted using a EUR loan, and received EUR collateral can be invested in the EUR account. Symmetric borrowing and investment are used here to preserve a linear pricing framework but, of course, the case of different borrowing and lending rates can also be covered by employing the theory of
backward stochastic differential equations.

\bd \label{def:ir-futures}
A \emph{SOFR futures} contract for the accrual period $[\Ud,\Td]$ is a futures contract referencing the
SOFR Average $\Rd(\Ud,\Td)$ and its quoted futures rate is denoted by $\Fd_t(\Ud,\Td)$ for
$t\in[0,\Td]$. An analogous \emph{\(\text{\euro STR}\) futures} contract is defined with the quoted futures rate $\Ff_t(\Ud,\Td)$.
\ed

We use a continuous-time variation-margin settlement, under which the futures positions have zero marked-to-market value and their gains are credited to the cash accounts. Initial margin and its funding are not modelled here. The dynamics of $\Fd$ and $\Ff$ will be specified later once we introduce an explicit term-structure model in Section~\ref{sec:cc-tsm}.

\subsection{Differential swaps and collateral conventions}\label{sec:swap-and-csa}

We work in a cross-currency setting and allow the collateral currency to be specified independently of the payoff currency. That is, the payoff currency and the collateral currency need not coincide.

We start by specifying a generic shape of the payoff of the swap contract. To keep the framework flexible, we introduce a one-parameter family indexed by a real constant $\gamma$, which scales the foreign overnight-rate leg. Varying
$\gamma$ allows the same notation to cover several distinct products within a unified pricing and hedging setup that enjoys the additivity property.

\bd\rm \label{def:sp-sofr-swap-fc}
A \textit{single-period SOFR/\(\text{\euro STR}\) swap} over $[\Ud,\Td]$, settled in arrears at $\Td$, is a contract settled in domestic currency with the following net payoff: at time $\Td$, the long party receives the floating amount  $\delta\Rd(\Ud,\Td)$ based on the domestic overnight rate and pays the floating amount $\delta\gamma\Rf(\Ud,\Td)$ based on the foreign overnight rate together with a fixed margin $\delta\kappa$.
Equivalently, the net cash flow at time $\Td$ for the long party is
\begin{equation}\label{eqlegs1}
\SwapT_\Td(\Ud,\Td)
= \delta\big(\Rd(\Ud,\Td)- \gamma \Rf(\Ud,\Td) -\kappa \big)\Noti,
\end{equation}
where $\Noti$ denotes the notional amount denominated in the domestic currency.
\ed

Let \(Q_t\) denote the spot FX rate, quoted as units of domestic currency per one unit of foreign currency.
A distinctive feature of the differential-swap specification is that the contractual payoff is \emph{domestic-currency settled} and does not involve \(Q\) explicitly. This differs from a cross-currency basis swap, where FX conversion introduces an explicit dependence on the exchange rate in the cash-flow specification; see \cite[Section~2.2]{Ding2024wp} for explicit definition and cash flows.

\brem \label{rem:swaps}
Definition~\ref{def:sp-sofr-swap-fc} and its multi-period extension are tailored to cover several instances of swap contracts that occur in market practice. When $\gamma=0$, the payoff given by \eqref{eqlegs1} reduces to a standard single-period SOFR swap: it is intrinsically domestic, although the associated CSA can specify either domestic or foreign collateralisation.
When $\gamma\neq 0$, the contract becomes a \emph{differential swap} driven by domestic and foreign overnight rates.
It is sometimes referred to as a \emph{quanto swap} when the constant $\gamma$ is interpreted as a predetermined conversion factor, for example, a fixed USD/EUR exchange rate. For further background on differential swaps in a LIBOR setting, see Wei \cite{Wei1994}.
\erem

We model collateral as variation margin posted or received over the life of the target trade. The CSA specifies its currency, the margin requirements and the remuneration rate. We compare a USD collateral specification with a EUR collateral specification.

We adopt the seller's replication convention: the hedger is short the target claim and holds a replicating portfolio whose net wealth $V$ equals the value of the claim to the long party. Let $i\in\{d,f\}$ denote the collateral currency specified by the CSA, with $d$ denoting USD and $f$ denoting EUR. Let $C_t^{\prime,i}$ denote the outstanding collateral principal at time $t$ measured in currency $i$, and let $C_t^i$ denote the USD value of that principal. In both symbols, the superscript $i$ identifies the collateral currency. The processes $C^d$ and $C^f$ correspond to the two alternative CSA specifications and need not coincide. We take $C_t^i>0$ for collateral received by the hedger and $C_t^i<0$ for collateral posted by the hedger. Throughout the paper, let $\beta=(\beta_t)_{t\geq0}$ be a deterministic process with values in $[0,\bar\beta]$, where $\bar\beta\geq1$ is a constant. Then the {\it proportional collateralisation} takes the form $C_t^i=-\beta_tV_t$, which means that a positive $V_t$ requires the hedger to post collateral. Full collateralisation corresponds to $\beta\equiv1$, whereas $\beta\equiv0$ gives the uncollateralised case. Values $\beta_t>1$, permitted when $\bar\beta>1$, represent overcollateralisation.

Write $Q_t^i$ for the USD value of one unit of currency $i$, so that $Q_t^d=1$ and $Q_t^f=Q_t$. Then
\[
C_t^i=Q_t^iC_t^{\prime,i}.
\]
Thus $C_t^d=C_t^{\prime,d}$ for USD collateral and $C_t^f=Q_tC_t^{\prime,f}$ for EUR collateral. Here, multiplication by $Q_t$ expresses the EUR principal in USD for valuation and does not entail an FX transaction. We leave the dependence of the strategy and wealth on the collateral currency implicit when it is clear from context. The collateral remuneration rate is
\[
r^{c,i}_t=r^i_t+\alpha^{c,i}_t
\]
where the spread $\alpha^{c,i}$ is deterministic and locally integrable. This corresponds to a CSA specifying the model's overnight benchmark together with a predetermined spread. The remuneration is paid in the collateral's native currency. Unless explicitly stated otherwise, we work under collateral conventions summarised in Assumption \ref{ass:collateral-policy}, which effectively describes a part of trading constraints for
self-financing futures strategies introduced in Section \ref{sec:collat-futures}.

\bhyp \label{ass:collateral-policy}
The CSA permits received cash collateral to be deposited in the money-market account of its native currency. The hedger adopts this investment policy and funds posted cash collateral by direct borrowing in that currency under Assumption~\ref{ass:direct-funding}. For EUR collateral received by the hedger, the outstanding principal is held in the EUR money-market account, while any (positive or negative) interest from the EUR account is immediately transferred to the USD account through spot conversion. Similarly, the EUR collateral posted by the hedger is funded from the EUR account and the negative interest cash flows in EUR are funded from the USD account through spot conversion. In addition, the collateral remuneration stipulated by the CSA is either paid or received in EUR and settled against the USD account through spot conversion. The EUR account is used only for the collateral deposit or the collateral funding loan in futures strategies introduced in Definition \ref{def:self-fin}.
\ehyp

The CSA and Assumption~\ref{ass:collateral-policy} have different roles: the former specifies the collateral obligations and permitted use, whereas the latter specifies how the hedger invests received collateral, funds posted collateral, and settles the interest cash flows. The received-side investment rate and the posted-side funding rate both equal $\bar r^i$ by Assumption~\ref{ass:direct-funding}. Counterparty default is not modelled here and thus partial collateralisation should be interpreted as an exogenous margin specification. Moreni and Pallavicini~\cite{MP2015} examine foreign-currency funding through collateralised FX swaps and derive pricing formulae for different currencies of contractual cash flows and collateral. For more information about general funding and collateral conventions, including borrowing and lending asymmetries, reuse and segregation, we refer to \cite{BR2015,BCR2018}.

Under the Gaussian term structure specification introduced later in Section~\ref{sec:cc-tsm}, the cross-currency expectation required for pricing the foreign floating leg is evaluated directly in Proposition~\ref{pro:single_sofr_foreign_collat} where it is shown to factorise into a domestic auxiliary discount factor, the traded \(\text{\euro STR}\) futures rate, and a deterministic quanto correction.

\subsection{Futures strategies for collateralised contracts}\label{sec:collat-futures}

Let $\Fd$ and $\Ff$ be continuous semimartingales representing the quoted SOFR and \(\text{\euro STR}\) futures rates. We first distinguish
the balances in the two cash accounts from the collateral return obligation. The process $C^i$ measures the collateral principal in USD, whereas $C^{\prime,i}$ measures it in the collateral currency. In particular, $C_t^f=Q_tC_t^{\prime,f}$ under EUR collateral and $C_t^d=C_t^{\prime,d}$ under USD collateral.

\bd  \label{def:fut-strat}
A \emph{futures trading strategy} $\Phi=(\phiv^{0,d},\phiv^{0,f},\phid,\phif)$ used to hedge the target trade consists of $\mathbb{F}$-predictable processes $\phid$ and $\phif$
representing positions  in SOFR and \(\text{\euro STR}\) futures, together with $\mathbb{F}$-adapted processes $\phiv^{0,d}$ and $\phiv^{0,f}$ representing holdings in the domestic and foreign money-market account, respectively. The values of the
domestic and foreign portfolios, measured in their own currencies, are
\begin{equation}\label{rty2}
V^{p,d}_t(\Phi,C^i)=\phiv^{0,d}_t\bar B^d_t,\qquad
V^{p,f}_t(\Phi,C^i)=\phiv^{0,f}_t\bar B^f_t,
\end{equation}
since the futures positions have zero marked-to-market value at any time. The portfolio's value and the hedger's net wealth, both measured in domestic currency, are, respectively,
\begin{equation}\label{eq:actual-book-and-net-wealth}
V^p_t(\Phi,C^i)=V^{p,d}_t(\Phi,C^i)+Q_tV^{p,f}_t(\Phi,C^i),
\qquad
V_t(\Phi,C^i)=V^p_t(\Phi,C^i)-C_t^i.
\end{equation}
\ed

When collateral is received, subtracting $C_t^i>0$ from the portfolio's value expressed in USD $V^p_t(\Phi,C^i)$ accounts for the obligation to return its principal at the contract's settlement date. When collateral is posted, adding $-C_t^i>0$ accounts for the receivable corresponding to its return from the counterparty. Under EUR collateral, a positive $V^{p,f}(\Phi,C^i)$ represents a deposit and a negative $V^{p,f}(\Phi,C^i)$ represents a loan used to fund the collateral
$C^{\prime,f}$ in its native currency. As expected, under USD collateral, no separate EUR cash balance is maintained and thus $\phiv^{0,f}$  vanishes. The arguments $(\Phi,C^i)$ record the dependence on the target trade's collateral process.

Consistently with Definition \ref{def:fut-strat}, for EUR collateral, we formally define the accumulated interest transfers from
the EUR account to domestic portfolio by the equalities
\begin{equation}\label{eq:collateral-interest-transfers}
D^{\prime,c}_t:=\int_0^t\bar r^f_uC_u^{\prime,f}\diff u,
\qquad
D^c_t:=\int_0^t Q_u\diff D^{\prime,c}_u=\int_0^t\bar r^f_uC_u^f\diff u.
\end{equation}
The process $D^{\prime,c}$ records interest earned on a deposit or interest payable on the EUR loan and the process
$D^c$ represents the flow of funds from the foreign account to (or from) the domestic portfolio.
Similarly, for either collateral currency, the accumulated remuneration cash flow stipulated by the CSA and measured in USD is given by
\begin{equation}\label{eq:collateral-remuneration-flow}
R^{c,i}_t:=-\int_0^t r^{c,i}_uC_u^i\diff u
\end{equation}
where the minus sign reflects the hedger's obligation to remunerate received collateral and entitlement to a positive interest on posted collateral. The conversion rule used to settle this payment is specified separately in Assumption~\ref{ass:collateral-policy}.

Foreign futures settlement is separate from the collateral currency. On a grid $0=t_0<\cdots<t_n=T$, the EUR variation-margin amount $\phif_{t_k}\Delta\Ff_{k+1}$ is converted into USD at the settlement-time rate $Q_{t_{k+1}}$, which yields
\begin{equation}\label{eq:right-end-futures-expansion}
Q_{t_{k+1}}\Delta\Ff_{k+1}
=Q_{t_k}\Delta\Ff_{k+1}
+\Delta Q_{k+1}\Delta\Ff_{k+1},
\end{equation}
and thus the continuous-time domestic-currency gains process for foreign futures is
\begin{equation}\label{fQsi}
\Ffq_t:=\Ffq_0+\int_0^t Q_u\diff\Ff_u+\qvar{Q,\Ff}{t}
\end{equation}
where the last two terms arise as the limits of the left-end sums and the products of increments, respectively; see \cite[Section~12.1]{oksendal2003sde}.

We are now ready to state a formal definition of a self-financing futures strategy with either collateralisation.
Notice that equations \eqref{eq:usd-account-bookkeeping}--\eqref{eq:eur-foreign-account-bookkeeping} are written from time zero before the first contractual payment. On each subsequent payment-free interval, the initial balances and integration limits are reset to the preceding payment date, and the accumulated interest transfers and remuneration are replaced by their increments over that interval.
All stochastic integrals and interest integrals appearing in Definition \ref{def:self-fin} are assumed well defined
and we also implicitly use equations \eqref{rty1} and \eqref{rty2}. For conciseness, we denote $V^{p,d}_t = V^{p,d}_t(\Phi,C^i)$
and $V^{p,f}_t=V^{p,f}_t(\Phi,C^i)$ in Definition \ref{def:self-fin}.

\bd \label{def:self-fin}
A futures strategy $\Phi$ is \emph{self-financing} under Assumptions~\ref{ass:direct-funding} and~\ref{ass:collateral-policy} and the preceding futures-settlement convention if, between contractual payment dates, its cash account values satisfy the following equations: \\
(i) for USD collateral, $V^{p,f}=0$ and
\begin{align} \label{eq:usd-account-bookkeeping}
V^{p,d}_t
={}V^{p,d}_0+\int_0^t\bar r^d_uV^{p,d}_u\diff u
+\int_0^t\phid_u\diff\Fd_u +\int_0^t\phif_uQ_u\diff\Ff_u
+\int_0^t\phif_u\diff\qvar{Q,\Ff}{u}
+R^{c,d}_t+C_t^d-C_0^d,
\end{align}
(ii) for EUR collateral, $V^{p,f}_0=C_0^{\prime,f}$ and
\begin{align}
V^{p,d}_t
={}&V^{p,d}_0+\int_0^t\bar r^d_uV^{p,d}_u\diff u
+\int_0^t\phid_u\diff\Fd_u +\int_0^t\phif_uQ_u\diff\Ff_u
+\int_0^t\phif_u\diff\qvar{Q,\Ff}{u}
+D^c_t+R^{c,f}_t,\label{eq:eur-domestic-account-bookkeeping}\\
V^{p,f}_t
={}&V^{p,f}_0+\int_0^t\bar r^f_uV^{p,f}_u\diff u
-D^{\prime,c}_t+C_t^{\prime,f}-C_0^{\prime,f}.\label{eq:eur-foreign-account-bookkeeping}
\end{align}
In case (i) we have $V^{p,d}_0=V_0+C_0^d$ and in case (ii) we have $V^{p,d}_0=V_0$ and in both cases the initial
net wealth $V_0$ is an arbitrary real number.
\ed

Under the present assumption of equal lending and borrowing rates, equation~\eqref{eq:eur-foreign-account-bookkeeping} covers both the EUR deposit and the EUR loan. Any increase or decrease in the collateral requirement changes the corresponding balance by the same amount and
interest on that balance is immediately either transferred to or funded from the USD account. These transfers of funds are formally represented
by the term $D^{\prime,c}$ in the EUR and the matching term $D^c$ in the USD in equations
\eqref{eq:eur-foreign-account-bookkeeping} and \eqref{eq:eur-domestic-account-bookkeeping}, respectively.
In contrast, in the case of USD collateral the principal is already encompassed by $V^{p,d}$ and its account interest is automatically
included in the  term $\bar r^dV^{p,d}$ so that the term $D^c$ is not present in \eqref{eq:usd-account-bookkeeping}.
Definition \ref{def:self-fin} combined with equation \eqref{eq:actual-book-and-net-wealth} lead to the following result on the
wealth dynamics under either a domestic or foreign collateralisation on intervals where no other incoming or outgoing contractual cash flows are made.

\bp \label{pro:collateral-account-reduction}
For a self-financing futures strategy $\Phi$ satisfying Definition~\ref{def:self-fin}, if collateral is received or posted in EUR, then
$V^{p,f}_t(\Phi,C^f)=C_t^{\prime,f}$ and $V_t(\Phi,C^f)=V^{p,d}_t(\Phi,C^f)$, whereas under USD collateralisation
we have $V_t(\Phi,C^d)=V^{p,d}_t(\Phi,C^d)-C_t^d$. In both cases, the net wealth satisfies the following equation on payment-free intervals, with the initial-date convention of Definition~\ref{def:self-fin}
\begin{align}\label{eq:continuous-self-fin}
V_t(\Phi,C^i)
={}&V_0(\Phi,C^i)+\int_0^t\bar r^d_uV_u(\Phi,C^i)\diff u
+\int_0^t(\bar r^i_u-r^{c,i}_u)C_u^i\diff u \nonumber\\
&+\int_0^t\phid_u\diff\Fd_u
+\int_0^t\phif_uQ_u\diff\Ff_u
+\int_0^t\phif_u\diff\qvar{Q,\Ff}{u}.
\end{align}
\ep

\begin{proof}
For EUR collateral, we subtract $C_t^{\prime,f}$ from~\eqref{eq:eur-foreign-account-bookkeeping} and use~\eqref{eq:collateral-interest-transfers} and $V^{p,f}_0(\Phi,C^f)=C_0^{\prime,f}$. The difference satisfies
\[
V^{p,f}_t(\Phi,C^f)-C_t^{\prime,f}
=\int_0^t\bar r^f_u\big(V^{p,f}_u(\Phi,C^f)-C_u^{\prime,f}\big)\diff u,
\]
and hence vanishes. The matched EUR account has USD value $Q_tV_t^{p,f}(\Phi,C^f)=C_t^f$, so that $V(\Phi,C^f)=V^{p,d}(\Phi,C^f)$ by~\eqref{eq:actual-book-and-net-wealth}. Substituting $D^c$ and $R^{c,f}$ into~\eqref{eq:eur-domestic-account-bookkeeping} gives~\eqref{eq:continuous-self-fin} with $i=f$. For USD collateral, we subtract $C_t^d$ from~\eqref{eq:usd-account-bookkeeping}, use the equality $V^{p,d}(\Phi,C^d)=V(\Phi,C^d)+C^d$, and substitute~\eqref{eq:collateral-remuneration-flow}.
This gives the same equation with $i=d$.
\end{proof}

Collateral principal transfers are recorded in their currency of denomination. Under EUR collateral, the matching account balance and the collateral return obligation or receivable offset each other and hence they cancel from the net wealth and its dynamics (though, of course, the interest payments and remuneration of collateral are still present in \eqref{eq:continuous-self-fin} since they affect the net wealth).

For brevity, we use the following reduced representation of $\Phi$ in the rest of the paper
\begin{equation}\label{eq:reduced-cash-position}
\phiv=(\phiv^0,\phid,\phif),
\qquad
\phiv^0_t:=(\bar B^d_t)^{-1} V_t(\Phi,C^i)
\end{equation}
where $\phiv^0$ is the net domestic cash position after accounting for the collateral principal. We write $V_t(\phiv,C^i):=V_t(\Phi,C^i)$ and $V^p_t(\phiv,C^i):=V^p_t(\Phi,C^i)$ for the wealth and portfolio value of the associated actual strategy. A reduced strategy is called self-financing when its associated actual strategy satisfies Definition~\ref{def:self-fin}. The actual money-market holdings
can be recovered from equalities
\begin{equation}\label{eq:actual-cash-holdings}
\phiv^{0,d}_t=(\bar B^d_t)^{-1} (V_t(\phiv,C^i)+\mathbf{1}_{\{i=d\}}C_t^i),
\qquad
\phiv^{0,f}_t=\mathbf{1}_{\{i=f\}}(\bar B^f_t)^{-1} C_t^{\prime,i}.
\end{equation}
Thus the EUR holding is prescribed by the collateral requirement rather than chosen as an additional hedge position. The restriction is on the replication strategies considered here; the cash funding facilities themselves are those of Assumption~\ref{ass:direct-funding}.

\brem \label{rem:collateral-conversion}
Let us make a few more comments on Assumption~\ref{ass:direct-funding} and its potential other variants.
Note that keeping received EUR principal in EUR and converting its interest into USD are decisions made at the discretion of the bank's treasury and are by no means necessary. Another possible choice would be to convert the principal into USD but this would have an effect
of replacing the EUR deposit by a USD asset while leaving the EUR return obligation unchanged and hence it would create an additional FX exposure that would need to be hedged. Likewise, retaining interest in EUR would leave a cash balance in addition to $C^{\prime,f}$ and produce a different FX exposure in the net wealth at the contract's settlement when the collateral amount is due to be returned to the pledgor. To avoid additional terms in a replicating strategy, we adopt the policy of Assumption~\ref{ass:collateral-policy} throughout but, in principle, alternative conventions and investment choices can also be dealt with using techniques developed in this work.
\erem

Along a replicating strategy, the process $V$ is the value of the target trade and, typically, the margin account
(variation of collateral) is based on the contract's mark-to-market value. This motivates the following commonly used
definition of a proportional collateralisation in which the function (or, simply, a constant) $\beta$ is specified by
the CSA.

\bhyp \label{ass:collateral-proportional}
The {\it proportional collateral rule} holds, that is, $C_t^i=-\beta_tV_t(\phiv,C^i)$ with the deterministic collateralisation process $\beta$ specified in Section~\ref{sec:swap-and-csa}.
\ehyp

Under Assumption \ref{ass:collateral-proportional}, which is adopted hereafter, Proposition~\ref{pro:collateral-account-reduction} yields
\begin{align}\label{eq:continuous-self-fin-effective}
\diff V_t(\phiv,C^i)
={}&r^{\beta,i}_tV_t(\phiv,C^i)\diff t
+\phid_t\diff\Fd_t
+\phif_tQ_t\diff\Ff_t
+\phif_t\diff\qvar{Q,\Ff}{t},\nonumber\\
r^{\beta,i}_t:={}&\bar r^d_t+\beta_t\big(r^{c,i}_t-\bar r^i_t\big),
\end{align}
and thus for domestic and foreign collateral we have
\begin{align}
r^{\beta,d}_t
&=r^d_t+\alpha^d_t
+\beta_t\big(\alpha^{c,d}_t-\alpha^d_t\big),
\label{eq:effective-rate-domestic}\\
r^{\beta,f}_t
&=r^d_t+\alpha^d_t
+\beta_t\big(\alpha^{c,f}_t-\alpha^f_t\big),
\label{eq:effective-rate-foreign}
\end{align}
respectively, which shows, in particular, that the benchmark rate $r^i$ cancels out within the collateral carry under EUR collateralisation.

Write $r^\beta=r^{\beta,i}$ for the effective rate of the target trade and define the \emph{effective account}
\[
B^\beta_t:=\exp\!\left(\int_0^t r^\beta_u\diff u\right),
\qquad B^\beta_0=1,
\]
which will be used as a discount factor for the net wealth. The process $B^\beta$ is introduced for mathematical
convenience but, obviously, it is not assumed to be a traded asset.

\bp\label{pro2.1}
Let $\phiv$ be a self-financing futures strategy in the sense of Definition~\ref{def:self-fin}.
Before the first contractual payment, the discounted net wealth $\wt V^\beta(\phiv,C^i)=(B^\beta)^{-1}V(\phiv,C^i)$ satisfies
\begin{align}
\wt V_t^\beta(\phiv,C^i)
={}&\wt V_0^\beta(\phiv,C^i)
+\int_0^t(B^\beta_u)^{-1}\phid_u\diff\Fd_u\nonumber\\
&+\int_0^t(B^\beta_u)^{-1}\phif_uQ_u\diff\Ff_u
+\int_0^t(B^\beta_u)^{-1}\phif_u\diff\qvar{Q,\Ff}{u}.
\end{align}
\ep

The effective rate combines the domestic funding rate and the net carry of the collateral deposit or funding loan in the selected currency.
Both receiving and posting lead to this rate because borrowing and investment use the same money market account.

\subsection{Pricing martingale measure}\label{def2.5}

Within the present linear setup, for either collateral currency, Proposition~\ref{pro2.1} reduces the collateralised pricing problem to the martingale properties of the two futures gains processes after discounting by the corresponding effective account. This motivates the following definition, which leads to the convenient pricing formula \eqref{eq:pricing}.

\bd \label{def:pmm}
A probability measure $\wtQ$ is called a \emph{pricing martingale measure (PMM)}
for the horizon $\Td$ if $\wtQ$ is equivalent to the statistical probability measure $\pp$ on $(\Omega,\cF_{\Td})$ and,
for either collateral currency, any collateralisation process $\beta$, and any futures strategy $\phiv$ that is self-financing for a target trade with no contractual payments before $\Td$ whose collateral process satisfies $C_t^i=-\beta_tV_t(\phiv,C^i)$, the process $(\wt V^\beta_t(\phiv,C^i))_{t\in[0,\Td]}$ is a $\wtQ$-local martingale.
\ed

Proposition~\ref{pro2.1} shows that the discounted wealth is a $\wtQ$-local martingale for every predictable choice of the futures positions for which the stochastic integrals are well defined if $\Fd$ and $\Ffq$ are $\wtQ$-local martingales. Thus a measure under which $\Fd$ and $\Ffq$ are local martingales can be used for either collateral currency and any proportional collateralisation process $\beta$. These choices change the effective account rather than the martingale condition on the futures gains processes. In Section~\ref{subsec:fut-pmm}, we specify a candidate measure and verify this property in Proposition~\ref{pro3.1}.

For a contract with a single terminal payoff, a self-financing strategy is admissible if its discounted net wealth $\wt V^\beta$ is a uniformly integrable $\wtQ$-martingale on $[0,\Td]$. For a contract with intermediate payments, the same condition is imposed on discounted ex-dividend wealth dynamically complemented by the accumulated discounted payments delivered by the strategy.

\bd\label{def:attainable}
A collateralised contract $(X_{\Td},\beta)$ with terminal payoff $X_{\Td}$ is \emph{attainable} if there exists an admissible futures strategy $\phiv$, which is self-financing for the contract with collateral process $C_t^i=-\beta_tV_t$, and such that $V_{\Td}(\phiv,C^i)=X_{\Td}$.
\ed

\bp \label{pro:pricing}
Assume that $(B^\beta_{\Td})^{-1}X_{\Td}$ is $\wtQ$-integrable and a collateralised contract $(X_{\Td},\beta)$ is attainable.
Then the arbitrage-free price process, which is equal to the value of its replicating strategy, satisfies, for every $t\in[0,\Td]$,
\begin{equation} \label{eq:pricing}
\pi_t^\beta(X_{\Td})=B^\beta_t\,\mathbb{E}_{\wtQ}\!\left[(B^\beta_{\Td})^{-1}X_{\Td}\,\big|\,\cF_t\right].
\end{equation}
\ep

\begin{proof}
Let $\phiv$ be an admissible futures strategy that replicates $X_{\Td}$, with associated collateral process $C^i$. By admissibility, its discounted wealth is a $\wtQ$-martingale, and attainability gives
\[
\wt V^\beta_{\Td}(\phiv,C^i)
=(B^\beta_{\Td})^{-1}X_{\Td}.
\]
Consequently,
\[
(B^\beta_t)^{-1}V_t(\phiv,C^i)
=\mathbb{E}_{\wtQ}\!\left[
(B^\beta_{\Td})^{-1}X_{\Td}\,\big|\,\cF_t
\right].
\]
Since the arbitrage-free value of an attainable claim equals the value of its replicating strategy, multiplication by $B^\beta_t$ yields \eqref{eq:pricing}.
\end{proof}

\brem\label{rem:pricing-representation-scope}
The pricing representation \eqref{eq:pricing} is not specific to futures-based replication. Its proof applies whenever the claim is replicated by a self-financing strategy whose discounted wealth is a uniformly integrable martingale under the corresponding pricing measure. Thus the same representation holds with other traded instruments under funding and collateral conventions yielding the same effective account; see also \cite{BR2015}. The available instruments determine attainability and the hedge positions, which we construct here using SOFR and \(\text{\euro STR}\) futures.
\erem

For brevity, we will sometimes write $\Xbeta_t$ instead of $\pi^\beta_t(X_{\Td})$ so that, in particular, $\Xbeta_{\Td}=X_{\Td}$.

\section{Cross-Currency Term Structure Model}\label{sec:cc-tsm}

This section introduces a tractable cross-currency term structure specification used to obtain explicit futures dynamics, hedge ratios, and numerical illustrations in later sections. Building on the model-free collateralised trading framework and the pricing martingale measure introduced in Section~\ref{sec:collat-funding}, we now impose a concrete Gaussian specification for the joint dynamics of the domestic and foreign overnight rates and the FX rate.

Section~\ref{subsec:standing} records a standing assumption that fixes the Vasicek--Garman--Kohlhagen dynamics of $(\rd,\rf,Q)$ under the domestic martingale measure $\Q$ and the corresponding dynamics under the foreign martingale measure $\whQ$. Section~\ref{subsec:aux} introduces auxiliary conditional discount factors under $\Q$ and $\whQ$ and summarises their diffusion coefficients, which serve as convenient inputs for futures dynamics and hedging ratios. Section~\ref{subsec:fut-pmm} states the diffusion structure of the traded SOFR and \(\text{\euro STR}\) futures and verifies that, within this specification, $\Q$ is a pricing martingale measure for the collateralised futures strategies. Finally, Section~\ref{subsec:restricted-complete} establishes a restricted completeness property for the class of claims considered in this work.

\subsection{Cross-currency model specification}\label{subsec:standing}

In a multi-curve cross-currency setting, the same economic model can be presented through several equivalent choices of numeraires and pricing measures. To avoid an extended discussion of measure construction, we record below a single standing assumption that fixes (i) the joint dynamics of the domestic and foreign short rates and the FX rate under a domestic pricing measure $\Q$, and (ii) the associated foreign martingale measure $\whQ$ obtained by an explicit Girsanov transform. All subsequent pricing and hedging arguments rely only on this specification. For background and derivations in the classical cross-currency setting, we refer to \cite[Chapter~14]{MR2005} and \cite{Ding2024wp}.

Recall from Section~\ref{sec:conventions} that $\xd=\rd+\alpha^d$ and $\xf=\rf+\alpha^f$ are the domestic and foreign money-market rates, respectively, and that the hedge funding rate is $r^h=\xd$. We set
\[
\lqt:=\alpha^d_t-\alpha^f_t,
\]
so that $\xd_t-\xf_t=\rd_t-\rf_t+\lqt$.

Under the domestic martingale measure, no arbitrage for the tradable money-market accounts of Assumption~\ref{ass:direct-funding} gives the FX relative drift $\bar r^d-\bar r^f$. Its spread relative to the difference of the overnight benchmarks is thus $\lambda^q=\alpha^d-\alpha^f$.

We assume that under $\Q$ the processes $\Zone,\Ztwo,\Zthree$ are one-dimensional Brownian motions with
instantaneous covariations
\[
\diff\qvar{Z^i,Z^j}{t}=\rho_{ij}\diff t.
\qquad i,j\in\{1,2,3\}.
\]
\bhyp\label{ass3.1}{\rm
Fix $\Td>0$ and let $\Q$ be a probability measure equivalent to $\pp$ on $(\Omega,\cF_{\Td})$.
Under the domestic pricing measure $\Q$, the domestic and foreign overnight rates $\rd,\rf$ follow Vasicek dynamics and the FX rate $Q$ (the domestic price to one unit of foreign currency) satisfies
\begin{align}
\diff\rd_t&=(a-b\rd_t)\diff t+\sigma\diff\Zone_t,  \nonumber\\
\diff\rf_t&=(\wh c-\wh b\,\rf_t)\diff t+\wh\sigma\diff\Ztwo_t,  \label{eq:Q-rates}\\
\diff Q_t&=Q_t(\rd_t-\rf_t+\lqt)\diff t+Q_t\barsig\diff\Zthree_t, \nonumber
\end{align}
where $a,b,\sigma,\wh c,\wh b,\wh\sigma,\barsig$ are positive constants and the processes $\Zone,\Ztwo$ and $\Zthree$ are one-dimensional Brownian motions under $\Q$ with the correlations $\diff \qvar{Z^i,Z^j}{t}=\rho_{ij} \diff t$. The correlation matrix is positive definite. The initial values $\rd_0,\rf_0$ and $Q_0$ are deterministic, with $Q_0>0$. Define the foreign martingale measure $\whQ$ on $(\Omega,\cF_{\Td})$ by the density
\begin{equation}\label{eq:foreign-density}
\frac{d\whQ}{d\Q}\Big|_{\cF_{\Td}}
:=\cE^q_{\Td},
\qquad
\cE^q_t:=\exp\!\Big(\barsig\,\Zthree_t-\frac12\barsig^2 t\Big),
\quad t\in[0,\Td].
\end{equation}
Then $\whZone_t=\Zone_t-\barsig\rhod\,t$, $\whZtwo_t=\Ztwo_t-\barsig\rhof\,t$ and
$\whZthree_t=\Zthree_t-\barsig\,t$ are correlated Brownian motions under $\whQ$.
Upon setting $R_t:=Q_t^{-1}$ and $\wh a:=\wh c+\wh\sigma\,\barsig\,\rhof$, the dynamics under $\whQ$ are
\begin{align*}
\diff\rd_t&=\big(a+\sigma\barsig\rhod-b\rd_t\big)\diff t+\sigma\diff\whZone_t,\\
\diff\rf_t&=(\wh a-\wh b\,\rf_t)\diff t+\wh\sigma\diff\whZtwo_t,\\
\diff R_t&=R_t(\rf_t-\rd_t-\lqt)\diff t-R_t\barsig\diff\whZthree_t.
\end{align*}
}\ehyp

\subsection{Dynamics of auxiliary discount factors}  \label{subsec:aux}

We introduce auxiliary conditional discount factors associated with the Vasicek specifications of the domestic and foreign overnight rates. These processes are \emph{not} postulated as additional traded securities. Rather, they provide a convenient way to isolate the diffusion terms that ultimately drive the dynamics of traded futures and, consequently, the hedge ratios derived in Section~\ref{sec:hedging}. We now recall standard affine computations for the Vasicek model \cite{Vasicek1977}.

We begin with the domestic auxiliary factor. For $u>0$ and $t\ge 0$ we set
\begin{equation}\label{eq:aux-Bd}
\Bd(t,u):=\EQ\!\left[\exp\!\Big(-\int_t^u \rd_v\diff v\Big)\,\Big|\,\cF_t\right].
\end{equation}
For $t\ge u$, the conditional expectation reduces to the accrual factor realised, that is,
$\Bd(t,u)=\exp(\int_u^t \rd_v\diff v)=(\Bd_u)^{-1}\Bd_t$, where $\diff \Bd_t=\rd_t \Bd_t\diff t$ and
$\Bd_0=1$. For $t\le u$, the standard arguments for the affine term structure model yield the semimartingale dynamics
\begin{equation}\label{eq:dyn-Bd}
\diff \Bd(t,u)=\Bd(t,u)\Big(\rd_t\diff t\,-\sigma n(t,u)\diff\Zone_t\Big),
\qquad t\in[0,u],
\end{equation}
where $n(t,u):=(1-e^{-b(u-t)})/b$.

We define the foreign auxiliary factor analogously, but under the foreign martingale measure $\whQ$.
For $u>0$ and $t\ge 0$ we set
\begin{equation}\label{eq:aux-Bf}
    \Bf(t,u):=\EwhQ \!\left[\exp\!\Big(-\int_t^u \rf_v\diff v\Big)\,\Big|\,\cF_t\right].
\end{equation}
For $t\le u$, under $\whQ$ we have the dynamics
\begin{equation}\label{eq:dyn-Bf}
    \diff \Bf(t,u)=\Bf(t,u)\Big(\rf_t\diff t-\wh \sigma \wh n (t,u)\diff\whZtwo_t\Big), \qquad t\in[0,u],
\end{equation}
where $\wh n(t,u):=(1-e^{-\wh b(u-t)})/\wh b$.

The following remark summarises the diffusion coefficients we will use later. We also record a simple
invariance principle that will be applied repeatedly: while drift terms change under equivalent measure
transformations, the continuous local martingale parts, and hence the diffusion coefficients relevant
for hedging remain unchanged. Specifically, for two continuous semimartingales $Y^1$ and $Y^2$ we write $Y^1\mart Y^2$ whenever $Y^1$ and $Y^2$ have the same continuous local martingale part in their canonical decompositions. If $Y^1\mart Y^2$ holds under $\Q$ on $(\Omega,\cF_T)$, then it also holds under any probability measure equivalent to $\Q$ on $(\Omega,\cF_T)$ and thus also under $\whQ$.

\brem\label{rem:aux-coeff}
For any fixed maturity $T>0$, the process $(\Bd(t,T),\,t\le T)$ satisfies
\[
\diff \Bd(t,T)\mart \Bd(t,T)\,\btT\diff\Zone_t,
\qquad \btT:=-\sigma\,n(t,T).
\]
Similarly, for any fixed maturity $T>0$, the process $(\Bf(t,T),\,t\le T)$ satisfies under $\whQ$
\[
\diff \Bf(t,T)\mart \Bf(t,T)\,\hbtT\diff\whZtwo_t,
\qquad \hbtT:=-\wh\sigma\,\wh n(t,T),
\]
and therefore $\diff \Bf(t,T)\mart \Bf(t,T)\,\hbtT\diff\Ztwo_t$ under $\Q$.
\erem

In Sections~\ref{subsec:fut-pmm} and \ref{sec:hedging} we will express the diffusion terms of the traded
SOFR and \(\text{\euro STR}\) futures, and the corresponding hedge ratios, directly in terms of
$\btT$ and $\hbtT$.

\subsection{Futures dynamics and the pricing martingale measure}\label{subsec:fut-pmm}

We now summarise the diffusion structure of the traded interest-rate futures that will be used for hedging.
Since explicit Vasicek-based representations are standard, we omit the derivations and refer to
\cite{Ding2024wp} for concise proofs and further details. Our emphasis here is on the semimartingale
decompositions and, in particular, on the diffusion coefficients that enter the hedge ratios.

Recall from Definition~\ref{def:benchmarks} that the backward-looking averages
$\Rd(\Ud,\Td)$ (SOFR) and $\Rf(\Ud,\Td)$ (\(\text{\euro STR}\)) are $\cF_{\Td}$-measurable. The corresponding
futures rates are defined as conditional expectations under the natural pricing measures: SOFR futures
are quoted/settled in the domestic currency and are therefore specified under $\Q$, while \(\text{\euro STR}\) futures are quoted and settled in the foreign currency and are specified under $\whQ$.

\bd\label{def:fut-ir}
The \emph{SOFR futures rate} $\Fd_t(\Ud,\Td)$ and the \emph{\(\text{\euro STR}\) futures rate} $\Ff_t(\Ud,\Td)$ are defined by
\[
\Fd_t(\Ud,\Td):=\EQ \big(\Rd(\Ud,\Td)\,\big|\,\cF_t\big),
\qquad
\Ff_t(\Ud,\Td):=\EwhQ \big(\Rf(\Ud,\Td)\,\big|\,\cF_t\big),
\qquad t\in[0,\Td].
\]
\ed

The next remark records the resulting diffusion coefficients. These coefficients will be reused later,
so we state them explicitly and adopt the shorthand notation introduced in
Section~\ref{subsec:aux}.

\brem\label{rem:fut-dyn}
Let $\Fd=\Fd(\Ud,\Td)$ and $\Ff=\Ff(\Ud,\Td)$ be the SOFR and \(\text{\euro STR}\) futures rates, respectively.
From \cite[Proposition~4.7]{Ding2024wp}, for \(t\in[0,\Td]\)
\[
\diff \Fd_t=\nud_t\diff\Zone_t,
\qquad
\nud_t
=\delta^{-1}\big(1+\delta\Fd_t(\Ud,\Td)\big)
\big(\mathbf{1}_{\{t\le\Ud\}}\btU-\btT\big).
\]
From \cite[Proposition~4.8]{Ding2024wp},
\[
\diff \Ff_t=\nuf_t\diff\whZtwo_t,
\qquad
\nuf_t
=\delta^{-1}\big(1+\delta\Ff_t(\Ud,\Td)\big)
\big(\mathbf{1}_{\{t\le\Ud\}}\hbtU-\hbtT\big),
\]
and hence $\diff \Ff_t\mart \nuf_t\diff\Ztwo_t$ under $\Q$.
Here $\delta=\Td-\Ud$.
\erem

Next, we reconcile these dynamics with the model-free pricing framework of Section~\ref{sec:collat-funding}.
Although collateralised valuation and hedging are most conveniently formulated under the pricing
martingale measure (PMM) $\wtQ$ introduced in Definition~\ref{def:pmm}, the term-structure model above
was specified under $\Q$. The following proposition verifies that, within the present setting, we may take $\wtQ=\Q$. The key step is to express the foreign futures gains in domestic currency and to
identify its diffusion term under $\Q$.

\bp\label{pro3.1}
The domestic martingale measure $\Q$ is a pricing martingale measure in the sense of Definition~\ref{def:pmm}. Hence we may take $\wtQ=\Q$.
\ep

\begin{proof}
Recall from \eqref{fQsi} that the domestic-currency gains process for foreign futures satisfies
\[
\Ffq_t=\Ffq_0+\int_0^t Q_u\diff \Ff_u+\qvar{Q,\Ff}{t}.
\]
Using Remark~\ref{rem:fut-dyn} and $\whZtwo=\Ztwo-\barsig\rhof\,t$, we obtain
\[
\diff \Ffq_t
=Q_t\nuf_t\big(\diff\whZtwo_t+\barsig\rhof\diff t\big)
=Q_t\nuf_t\diff\Ztwo_t.
\]
Hence $\Ffq$ is a (local) martingale under $\Q$. Since $\Fd$ is also a $\Q$-martingale by definition, $\Q$
is a pricing martingale measure for the traded futures family considered in this paper, under either collateral currency and for every admissible proportional collateralisation process \(\beta\).
\end{proof}

The next remark records the resulting diffusion coefficient for the domestic-currency foreign-futures
process, which will be used repeatedly when expressing hedge ratios in Section~\ref{sec:hedging}.

\brem\label{rem:Ffq-dyn}
In particular, $\diff \Ffq_t=\nufq_t\diff\Ztwo_t$ where $\nufq_t:=Q_t\nuf_t$ and $\nuf_t$ is given in
Remark~\ref{rem:fut-dyn}.
\erem
\subsection{Restricted completeness}\label{subsec:restricted-complete}
The three-factor specification we introduced in Assumption \ref{ass3.1} implies that the market is incomplete for general $\cF_{\Td}$-measurable claims when trading strategies are restricted to $(\Fd,\Ff)$. Nevertheless, the products considered in this paper share a structural feature: their discounted price processes do not load on the third Brownian component. Consequently, they are hedgeable using $(\Fd,\Ffq)$ and thus $(\Fd,\Ff)$ only. Let $\ff^{(2)}:=(\cF_t^{\Zone,\Ztwo})_{t\in[0,\Td]}$ be the subfiltration generated by $(\Zone,\Ztwo)$. Here, $\cF_t^{\Zone,\Ztwo}:=\sigma\!\big(\Zone_u,\Ztwo_u:\,0\le u\le t\big)\vee\mathcal{N}$ denotes the usual augmentation, where $\mathcal{N}$ denotes the collection of $\pp$-null sets in $\cF$.
Hence $(\Zone,\Ztwo)$ enjoys the predictable representation property with respect to $\ff^{(2)}$.

\bp\label{pro:restricted-complete}
Consider a collateralised contract $(X_{\Td},\beta)$ with terminal payoff $X_{\Td}$ at time $\Td$ and
proportional collateralisation at rate $\beta$. Assume that $X_{\Td}(B^{\beta}_{\Td})^{-1}$ is
$\Q$-integrable and that the martingale
\[
M_t:=\EQ\!\left[(B^\beta_{\Td})^{-1}X_{\Td}\,\big|\,\cF_t\right],
\qquad t\in[0,\Td],
\]
is $\ff^{(2)}$-adapted. Then $(X_{\Td},\beta)$ can be replicated by an admissible self-financing strategy with net wealth $V_t=B^\beta_tM_t$. Its futures positions are
\[
\phid_t=\frac{\psi^1_t}{\nud_t},
\qquad
\phif_t=\frac{\psi^2_t}{\nufq_t},
\qquad 0\le t<\Td,
\]
where $(\psi^1,\psi^2)$ is $\ff^{(2)}$-predictable and satisfies
\[
\diff M_t=(B^\beta_t)^{-1}\big(\psi^1_t\diff\Zone_t+\psi^2_t\diff\Ztwo_t\big).
\]
The futures positions are $\mathbb F$-predictable and unique up to $dt\otimes\Q$-null sets before expiry. The actual cash holdings are recovered from~\eqref{eq:actual-cash-holdings}.
\ep

\begin{proof}
The martingale $M$ is uniformly integrable. Since $\beta$ and the spreads are deterministic, the effective account $B^\beta$ is continuous and $\ff^{(2)}$-adapted. The predictable representation property of $(\Zone,\Ztwo)$ therefore gives the stated representation with coefficients $(\psi^1,\psi^2)$. By Remarks~\ref{rem:fut-dyn} and~\ref{rem:Ffq-dyn},
\[
\diff\Fd_t=\nud_t\diff\Zone_t,\qquad
\diff\Ffq_t=\nufq_t\diff\Ztwo_t.
\]
The coefficients $\nud_t$ and $\nufq_t$ are strictly positive for $t<\Td$ and vanish at $\Td$. Substituting the stated positions gives $\diff\wt V^\beta=\diff M$ with $\wt V^\beta_0=M_0$, and hence $V=B^\beta M$ and $V_{\Td}=X_{\Td}$. Uniform integrability of $M$ establishes admissibility. Positive definiteness of the correlation matrix gives uniqueness of the futures positions up to $dt\otimes\Q$-null sets.
\end{proof}

\brem\label{rem:restricted}
Adaptedness of $M$ to $\ff^{(2)}$ is a sufficient condition for replication in Proposition~\ref{pro:restricted-complete}. The discounted payoffs considered in Section~\ref{sec:hedging} satisfy this condition.

Moreover, suppose that $\gamma=0$. The payoff then depends only on the domestic overnight rate, while equations~\eqref{eq:effective-rate-domestic}--\eqref{eq:effective-rate-foreign} show that, under either collateral currency, the effective rate contains no foreign-rate factor. Consequently, the discounted price process has no exposure to $\Ztwo$. Since there is no foreign-rate risk to hedge, its martingale coefficient satisfies $\psi^2=0$ and Proposition~\ref{pro:restricted-complete} gives $\phif=0$.
\erem

\section{Arbitrage-Free Pricing under Collateral Currency Choice}\label{sec:pricing}

This section derives arbitrage-free prices for USD-settled SOFR/\(\text{\euro STR}\) differential swaps under proportional collateralisation in either USD or EUR. Proposition~\ref{pro3.1} identifies the pricing martingale measure \(\wtQ\) with the domestic martingale measure \(\Q\), so all conditional expectations below are taken under this measure.

Recall that, for collateral currency \(i\), we write
\[
\bar r^i=r^i+\alpha^i,
\qquad
r^{c,i}=r^i+\alpha^{c,i},
\]
where \(\bar r^i\) is the rate earned on the retained collateral principal and \(r^{c,i}\) is the collateral remuneration rate. The spreads \(\alpha^d,\alpha^i,\alpha^{c,i}\) are deterministic. The bookkeeping of Section~\ref{sec:collat-futures} gives
\begin{equation}\label{eq:effective-rate-general}
r^{\beta,i}_t
=\bar r^d_t+\beta_t\big(r^{c,i}_t-\bar r^i_t\big)
=r^d_t+\alpha^{\beta,i}_t,
\qquad
\alpha^{\beta,i}_t
:=\alpha^d_t+\beta_t\big(\alpha^{c,i}_t-\alpha^i_t\big).
\end{equation}
Accordingly, for \(t\le S\), define
\begin{equation}\label{eq:A-beta-foreign}
A^{\beta,i}_{t,S}
:=\exp\!\left(-\int_t^S\alpha^{\beta,i}_u\diff u\right).
\end{equation}
For every attainable payoff \(X_S\),
\begin{equation}\label{eq:pricing-domestic-factor}
\pi^{\beta,i}_t(X_S)
=A^{\beta,i}_{t,S}
\EQ\!\left[
e^{-\int_t^S r^d_u\diff u}X_S
\,\Big|\,\cF_t\right].
\end{equation}
Thus the collateral currency enters through the deterministic adjustment \(A^{\beta,i}_{t,S}\) to domestic discounting, whereas the contractual cash-flow currency and the reference rates determine \(X_S\).

A multi-period differential swap is the sum of its coupon claims. Since the pricing operator is linear, its valuation reduces to the corresponding single-period problem: we first price one coupon and then sum the resulting values over the outstanding payment dates. We first compute the conditional-expectation values and their attainability is established in Section~\ref{sec:hedging}.

\subsection{Single-period differential swaps under collateralisation}
\label{sec:partial-foreign-coll}\label{sec:full-foreign-coll}

We consider the single-period swap of Definition~\ref{def:sp-sofr-swap-fc} over \([\Ud,\Td]\), settled at \(\Td\), and set \(\Noti=1\). Using
\[
\delta\Rd(\Ud,\Td)=e^{\int_{\Ud}^{\Td}\rd_u\diff u}-1,
\qquad
\delta\Rf(\Ud,\Td)=e^{\int_{\Ud}^{\Td}\rf_u\diff u}-1,
\]
its terminal cash flow at time \(\Td\) is
\begin{equation}\label{eqlegs}
\SwapT_{\Td}(\Ud,\Td)
=e^{\int_{\Ud}^{\Td}\rd_u\diff u}
-\gamma e^{\int_{\Ud}^{\Td}\rf_u\diff u}
-(1-\gamma+\delta\kappa).
\end{equation}

\bp\label{pro:single_sofr_foreign_collat}
\label{pro:sp-sofr-full-foreign-coll}
\label{pro:sp-sofr-full-foreign-coll-two}
For every \(t\in[0,\Td]\), the arbitrage-free price under proportional collateralisation in currency \(i\) is
\begin{equation}\label{eq:single-price-new}
\Swapb_t(\Ud,\Td;i)
:=\pi^{\beta,i}_t\big(\SwapT_{\Td}(\Ud,\Td)\big)
=I^{(1)}_t-\gamma I^{(2)}_t-(1-\gamma+\delta\kappa)I^{(3)}_t,
\end{equation}
where
\begin{equation}\label{eq:I1-new}
I^{(1)}_t
=A^{\beta,i}_{t,\Td}
\begin{cases}
\Bd(t,\Ud),&0\le t\le\Ud,\\[1mm]
(B^d_{\Ud})^{-1}B^d_t,&\Ud\le t\le\Td,
\end{cases}
\end{equation}
\begin{equation}\label{eq:I2-new}
I^{(2)}_t
=A^{\beta,i}_{t,\Td}\Bd(t,\Td)
\big(1+\delta\Ff_t(\Ud,\Td)\big)
\Gamma^{\mathrm{qf}}_t(\Ud,\Td),
\end{equation}
and
\begin{equation}\label{eq:I3-new}
I^{(3)}_t
=A^{\beta,i}_{t,\Td}\Bd(t,\Td).
\end{equation}
Here, with
\[
\sigma_S(u):=-\sigma n(u,S),
\qquad
\wh{\sigma}_S(u):=-\wh{\sigma}\,\wh n(u,S),
\]
the deterministic quanto correction term is
\begin{equation}\label{eq:quanto-correction}
\Gamma^{\mathrm{qf}}_t(\Ud,\Td)
:=
\begin{cases}
\displaystyle
\exp\!\left(
\int_t^{\Ud}
\big(\rhodf\sigma_{\Td}(u)-\rhof\barsig\big)
\wh{\sigma}_{\Ud}(u)\diff u
-\int_t^{\Td}
\big(\rhodf\sigma_{\Td}(u)-\rhof\barsig\big)
\wh{\sigma}_{\Td}(u)\diff u
\right),&t\le\Ud,\\[3mm]
\displaystyle
\exp\!\left(
-\int_t^{\Td}
\big(\rhodf\sigma_{\Td}(u)-\rhof\barsig\big)
\wh{\sigma}_{\Td}(u)\diff u
\right),&t\ge\Ud.
\end{cases}
\end{equation}
\ep

\begin{proof}
Equation~\eqref{eq:pricing-domestic-factor} and linearity give~\eqref{eq:single-price-new}. For the domestic floating component,
\[
I^{(1)}_t
=A^{\beta,i}_{t,\Td}
\EQ\!\left[
e^{-\int_t^{\Td}\rd_u\diff u}
e^{\int_{\Ud}^{\Td}\rd_u\diff u}
\,\Big|\,\cF_t\right].
\]
If \(t\le\Ud\), the two exponentials reduce to \(e^{-\int_t^{\Ud}\rd_u\diff u}\); if \(t\ge\Ud\), they reduce to the \(\cF_t\)-measurable factor
\(e^{\int_{\Ud}^{t}\rd_u\diff u}=(B^d_{\Ud})^{-1}B^d_t\).
This proves~\eqref{eq:I1-new}, while~\eqref{eq:I3-new} follows directly from the definition of \(\Bd(t,\Td)\).

\noindent For the foreign floating component, conditionally on \(\cF_t\), the random variables \(-\int_t^{\Td}\rd_u\diff u,
\int_{\Ud}^{\Td}\rf_u\diff u,\) and \(
\log\!\left(\frac{\cE^q_{\Td}}{\cE^q_t}\right)\)
are jointly Gaussian. Moreover,
\[
1+\delta\Ff_t(\Ud,\Td)
=
\EwhQ\!\left[
\exp\!\left(\int_{\Ud}^{\Td}\rf_u\diff u\right)
\,\Big|\,\cF_t
\right]
=
\EQ\!\left[
\frac{\cE^q_{\Td}}{\cE^q_t}
\exp\!\left(\int_{\Ud}^{\Td}\rf_u\diff u\right)
\,\Big|\,\cF_t
\right].
\]
The conditional Gaussian exponential-moment formula therefore gives
\begin{align}
&\EQ\!\left[
\exp\!\left(
-\int_t^{\Td}\rd_u\diff u
+\int_{\Ud}^{\Td}\rf_u\diff u
\right)
\,\Big|\,\cF_t
\right] \nonumber\\
&\quad=
\EQ\!\left[
\exp\!\left(-\int_t^{\Td}\rd_u\diff u\right)
\,\Big|\,\cF_t
\right]
\EwhQ\!\left[
\exp\!\left(\int_{\Ud}^{\Td}\rf_u\diff u\right)
\,\Big|\,\cF_t
\right] \nonumber\\
&\qquad\quad\times
\exp\!\left(
\operatorname{Cov}_t\!\left(
-\int_t^{\Td}\rd_u\diff u,
\int_{\Ud}^{\Td}\rf_u\diff u
\right)
-
\operatorname{Cov}_t\!\left(
\log\!\left(\frac{\cE^q_{\Td}}{\cE^q_t}\right),
\int_{\Ud}^{\Td}\rf_u\diff u
\right)
\right).
\label{eq:quanto-foreign-leg}
\end{align}
For \(t\le\Ud\), the Brownian loading of
\(\int_{\Ud}^{\Td}\rf_u\diff u\) is
\(\wh{\sigma}_{\Ud}(u)-\wh{\sigma}_{\Td}(u)\) on \([t,\Ud]\) and
\(-\wh{\sigma}_{\Td}(u)\) on \([\Ud,\Td]\). The Brownian loadings of
\(-\int_t^{\Td}\rd_u\diff u\) and
\(\log(\cE^q_{\Td}/\cE^q_t)\) are
\(\sigma_{\Td}(u)\) and \(\barsig\), respectively.
Substitution into~\eqref{eq:quanto-foreign-leg} gives the first line
of~\eqref{eq:quanto-correction}. For \(t\ge\Ud\), the Brownian loading of
\(\int_{\Ud}^{\Td}\rf_u\diff u\) over \([t,\Td]\) is
\(-\wh{\sigma}_{\Td}(u)\), which gives the second line
of~\eqref{eq:quanto-correction}. Finally, by definition of \(\Bd(t,\Td)\), equation~\eqref{eq:I2-new} follows.
\end{proof}

\brem\label{rem:two-representations-hedging}
For USD collateral,
\[
\alpha^{\beta,d}_t
=(1-\beta_t)\alpha^d_t+\beta_t\alpha^{c,d}_t,
\]
so the effective spread interpolates between the domestic funding spread and the domestic collateral-remuneration spread. For EUR collateral,
\[
\alpha^{\beta,f}_t
=\alpha^d_t+\beta_t(\alpha^{c,f}_t-\alpha^f_t),
\]
so the adjustment is the net EUR collateral carry added to the domestic funding spread. Full EUR collateralisation corresponds to $\beta\equiv1$.

If \(\gamma=0\), the foreign component \(I^{(2)}\) is absent, so neither \(r^f\) nor the quanto correction \(\Gamma^{\mathrm{qf}}\) enters the price. Since all three components in~\eqref{eq:single-price-new} share the same factor \(A^{\beta,i}_{t,\Td}\), the fair fixed margin of a single-period swap is independent of both \(\beta\) and \(i\).
\erem

\subsection{Extension to multi-period differential swaps}\label{sec:multi-period-foreign-coll}
We now extend the single-period result by applying Proposition~\ref{pro:single_sofr_foreign_collat} coupon by coupon and summing the resulting ex-dividend values.
Let \(0\le T_0<T_1<\cdots<T_n\) be the payment schedule and set
\(\delta_j:=T_j-T_{j-1}\).
The \(j\)th coupon, paid at \(T_j\), equals
\[
e^{\int_{T_{j-1}}^{T_j}\rd_u\diff u}
-\gamma e^{\int_{T_{j-1}}^{T_j}\rf_u\diff u}
-(1-\gamma+\delta_j\kappa).
\]

\bp\label{pro:mp-sofr-foreign-collat}
For \(t\in[0,T_n]\), let
\(k(t):=\min\{j\in\{1,\ldots,n\}:t<T_j\}\) for \(t<T_n\), and set \(k(T_n):=n+1\). Empty sums are zero.
The ex-dividend price under proportional collateralisation in currency \(i\) is
\begin{align}\label{eq:multi-price-new}
\Swapb_t(T_0,n;i)
&=\sum_{j=k(t)}^n
\Big(I^{(1)}_j(t)-\gamma I^{(2)}_j(t)
-(1-\gamma+\delta_j\kappa)I^{(3)}_j(t)\Big),
\end{align}
where
\begin{equation}\label{eq:multi-I1-new}
I^{(1)}_j(t)
=A^{\beta,i}_{t,T_j}
\begin{cases}
\Bd(t,T_{j-1}),&t\le T_{j-1},\\[1mm]
(B^d_{T_{j-1}})^{-1}B^d_t,&T_{j-1}\le t\le T_j,
\end{cases}
\end{equation}
\begin{equation}\label{eq:multi-I2-new}
I^{(2)}_j(t)
=A^{\beta,i}_{t,T_j}\Bd(t,T_j)
\big(1+\delta_j\Ff_t(T_{j-1},T_j)\big)
\Gamma^{\mathrm{qf}}_t(T_{j-1},T_j),
\end{equation}
and
\begin{equation}\label{eq:multi-I3-new}
I^{(3)}_j(t)
=A^{\beta,i}_{t,T_j}\Bd(t,T_j).
\end{equation}
The price is measured after any payment at time $t$. At each payment date $T_j$,
\[
\Swapb_{T_j-}(T_0,n;i)-\Swapb_{T_j}(T_0,n;i)
=\SwapT_{T_j}(T_{j-1},T_j).
\]
In particular, $\Swapb_{T_n}(T_0,n;i)=0$ after delivery of the last coupon.
\ep

\begin{proof}
The swap is the sum of its remaining coupons and the pricing operator is linear. Applying Proposition~\ref{pro:single_sofr_foreign_collat} to each coupon with
\((\Ud,\Td,\delta)\) replaced by
\((T_{j-1},T_j,\delta_j)\) gives~\eqref{eq:multi-price-new}--\eqref{eq:multi-I3-new}.
\end{proof}
For a newly initiated swap, the fixed margin is quoted so that the swap has zero value at the quoting time. Solving the multi-period price for this margin gives the corresponding par fixed margin.
\bcor\label{cor:par-rate-mp-sofr-foreign-collat}
For \(t\in[0,T_0]\), the unique par fixed margin is
\begin{equation}\label{eq:par-swap-rate-mp-diff}
\kappa^\star_t(T_0,n;\beta,\gamma,i)
=\frac{\displaystyle
\sum_{j=1}^n\Big(I^{(1)}_j(t)-\gamma I^{(2)}_j(t)
-(1-\gamma)I^{(3)}_j(t)\Big)}
{\displaystyle\sum_{j=1}^n\delta_j I^{(3)}_j(t)}.
\end{equation}
Equivalently,
\begin{equation}\label{eq:swap-par-representation-mp-diff}
\Swapb_t(T_0,n;i)
=\big(\kappa^\star_t(T_0,n;\beta,\gamma,i)-\kappa\big)
\sum_{j=1}^n\delta_j I^{(3)}_j(t).
\end{equation}
\ecor

\brem\label{rem:multi-beta-special-cases}
Unlike the single-period case, the multi-period par margin can depend on both \(\beta\) and the collateral currency \(i\), because the deterministic factors \(A^{\beta,i}_{t,T_j}\) reweight cash flows paid at different dates. If a CSA admits a set \(\mathcal C\) of eligible collateral currencies and the selected currency is retained over the life of the trade, the model gives a static comparison through the currency-specific net carry spreads
\[
\alpha^{c,i}-\alpha^i,
\qquad i\in\mathcal C.
\]
Within the collateral carry \(r^{c,i}-\bar r^i\), the benchmark rate \(r^i\) cancels between the return earned on the collateral principal and the collateral remuneration. The resulting fixed-currency ranking is therefore governed by the net carry spreads. A dynamic cheapest-to-deliver problem, as studied for example in \cite{Fujii2011Choice,Wolf2022}, additionally requires a right to substitute collateral. Stochastic relative carry spreads can generate collateral-switching convexity, which is outside the present fixed-currency framework.

When \(\gamma=0\), equations~\eqref{eq:multi-I1-new} and~\eqref{eq:multi-I3-new} show that the price is driven only by the domestic rate factor. The chosen collateral currency affects discounting through the deterministic factors \(A^{\beta,i}_{t,T_j}\) and creates no foreign-rate risk. Consequently, the position in \(\text{\euro STR}\) futures is zero, as shown later in Proposition~\ref{pro:hedge-mp-collateral-choice}.
\erem
\section{Futures-Based Hedging under Collateral Currency Choice}\label{sec:hedging}
Having derived the arbitrage-free price processes in Section~\ref{sec:pricing}, we now verify their attainability by constructing self-financing strategies in the traded futures. Since the currency-specific adjustment \(A^{\beta,i}\) is deterministic, the collateral specification changes the scale of the price components but does not introduce an additional Brownian risk factor. The foreign-rate exposure is generated by the \(\text{\euro STR}\) leg of the contractual payoff when \(\gamma\ne0\).

For the futures associated with $[\Ud,\Td]$, suppressing their accrual dates, the self-financing wealth dynamics are
\begin{equation}\label{eq:wealth-hedging-general}
\diff V_t=r^{\beta,i}_tV_t\diff t+\phid_t\diff\Fd_t+\phif_tQ_t\diff\Ff_t+\phif_t\diff\qvar{Q,\Ff}{t}.
\end{equation}

where \(r^{\beta,i}\) is given by~\eqref{eq:effective-rate-general}. The domestic-currency foreign-futures gains process satisfies
\(\diff\Ffq_t=\nufq_t\diff\Ztwo_t\), as stated in Remark~\ref{rem:Ffq-dyn}.

The net domestic cash position is the reduced position of~\eqref{eq:reduced-cash-position}, so we have \(\phiv^0_t=\frac{V_t}{\bar B^d_t}\). The actual bank holdings are given by~\eqref{eq:actual-cash-holdings}. With $C_t^i=-\beta_tV_t$, the portfolio value is $V^p_t=V_t+C_t^i=(1-\beta_t)V_t$. When $\beta\equiv1$, it vanishes under either collateral currency. Under USD collateral the actual USD account balance is then zero. Under EUR collateral the USD balance $V^{p,d}=V$ is offset by the EUR account balance, whose domestic value is $QV^{p,f}=C^f=-V$.

The multi-period swap is the sum of its outstanding coupon claims. Since both the pricing operator and the self-financing gains equation are linear, it is enough to construct the hedge for a single coupon; Proposition~\ref{pro:hedge-mp-collateral-choice} then obtains the multi-period strategy by summing the coupon-level positions.
\subsection{Hedging of single-period differential swaps}
\label{sec:hedge-sofr-partial-foreign}
\label{sec:hedge-sofr-full-foreign}

Let \(I^{(1)},I^{(2)},I^{(3)}\) be the pricing components in Proposition~\ref{pro:single_sofr_foreign_collat}, so that
\[
\Swapb_t(\Ud,\Td;i)
=I^{(1)}_t-\gamma I^{(2)}_t
-(1-\gamma+\delta\kappa)I^{(3)}_t.
\]

\bp\label{pro:hedge-sp-sofr-partial-foreign}
\label{pro:hedge-sp-sofr-full-foreign}
The single-period differential swap with terminal payoff
\(\SwapT_{\Td}(\Ud,\Td)\) and proportional collateralisation in currency \(i\) can be replicated by a self-financing futures strategy
\[
\phiv=(\phiv^0,\phid,\phif),
\]
with associated target-trade collateral process $C^i$, such that
\[
V_t(\phiv,C^i)=\Swapb_t(\Ud,\Td;i),
\qquad
C_t^i=-\beta_t\Swapb_t(\Ud,\Td;i).
\]
The net domestic cash position is
\begin{equation}\label{eq:phi0-collateral-choice}
\phiv^0_t
=\frac{\Swapb_t(\Ud,\Td;i)}{\bar B^d_t}.
\end{equation}

The actual money-market holdings corresponding to this reduced strategy are

\begin{equation}\label{eq:actual-single-period-cash-holdings}
\begin{aligned}
\phiv^{0,d}_t
=\frac{\big(1-\beta_t\mathbf{1}_{\{i=d\}}\big)\Swapb_t(\Ud,\Td;i)}{\bar B^d_t},
\qquad\phiv^{0,f}_t
=\mathbf{1}_{\{i=f\}}\frac{C_t^{\prime,i}}{\bar B^f_t}.
\end{aligned}
\end{equation}
For \(0\le t<\Td\), the futures positions are as follows.
\begin{equation}\label{eq:phi-d-collateral-choice}
\phid_t
=
\frac{
\mathbf{1}_{\{t\le\Ud\}}\btU I^{(1)}_t
-\btT\big(
\gamma I^{(2)}_t
+(1-\gamma+\delta\kappa)I^{(3)}_t
\big)}
{\nud_t}.
\end{equation}
In particular, for \(t\in[\Ud,\Td)\),
\begin{equation}\label{eq:phi-d-collateral-choice-accrual}
\phid_t
=
\frac{
\delta\big(
\gamma I^{(2)}_t
+(1-\gamma+\delta\kappa)I^{(3)}_t
\big)}
{1+\delta\Fd_t(\Ud,\Td)}.
\end{equation}
The \(\text{\euro STR}\)-futures position is
\begin{align}
\phif_t
&=
-\frac{
\gamma\big(
\mathbf{1}_{\{t\le\Ud\}}\hbtU-\hbtT
\big)I^{(2)}_t}
{\nufq_t} \nonumber\\
&=
-\frac{\gamma\delta I^{(2)}_t}
{Q_t\big(1+\delta\Ff_t(\Ud,\Td)\big)}
=
-\frac{\gamma\delta}{Q_t}
A^{\beta,i}_{t,\Td}\Bd(t,\Td)
\Gamma^{\mathrm{qf}}_t(\Ud,\Td).
\label{eq:phi-f-collateral-choice}
\end{align}
\ep

\begin{proof}
The factors \(A^{\beta,i}\) and \(\Gamma^{\mathrm{qf}}\) are deterministic. Using Proposition~\ref{pro:single_sofr_foreign_collat} together with Remarks~\ref{rem:aux-coeff} and~\ref{rem:fut-dyn}, the local martingale parts of the three pricing components are
\begin{align}
\diff I^{(1)}_t
&\mart
\mathbf{1}_{\{t\le\Ud\}}\btU I^{(1)}_t\diff\Zone_t,
\label{eq:dI1-hedging-new}\\
\diff I^{(2)}_t
&\mart
\btT I^{(2)}_t\diff\Zone_t
+\big(
\mathbf{1}_{\{t\le\Ud\}}\hbtU-\hbtT
\big)I^{(2)}_t\diff\Ztwo_t,
\label{eq:dI2-hedging-new}\\
\diff I^{(3)}_t
&\mart
\btT I^{(3)}_t\diff\Zone_t.
\label{eq:dI3-hedging-new}
\end{align}
Indeed, the \(\diff\Ztwo\)-coefficient in~\eqref{eq:dI2-hedging-new} follows from
\[
\frac{\delta\nuf_t}
{1+\delta\Ff_t(\Ud,\Td)}
=
\mathbf{1}_{\{t\le\Ud\}}\hbtU-\hbtT.
\]
Consequently,
\begin{align}
\diff\Swapb_t(\Ud,\Td;i)
\mart{}&
\Big[
\mathbf{1}_{\{t\le\Ud\}}\btU I^{(1)}_t
-\btT\big(
\gamma I^{(2)}_t
+(1-\gamma+\delta\kappa)I^{(3)}_t
\big)
\Big]\diff\Zone_t \nonumber\\
&\quad
-\gamma\big(
\mathbf{1}_{\{t\le\Ud\}}\hbtU-\hbtT
\big)I^{(2)}_t\diff\Ztwo_t.
\label{eq:swap-martingale-hedging-new}
\end{align}
By Remarks~\ref{rem:fut-dyn} and~\ref{rem:Ffq-dyn},
\[
\diff\Fd_t=\nud_t\diff\Zone_t,
\qquad
\diff\Ffq_t=\nufq_t\diff\Ztwo_t.
\]
Matching the two diffusion coefficients in~\eqref{eq:swap-martingale-hedging-new} gives~\eqref{eq:phi-d-collateral-choice} and the first equality in~\eqref{eq:phi-f-collateral-choice}. The remaining equalities in~\eqref{eq:phi-f-collateral-choice} follow from
\(\nufq=Q\nuf\) and equation~\eqref{eq:I2-new}. Equation~\eqref{eq:phi-d-collateral-choice-accrual} follows from the expression for \(\nud\) on \([\Ud,\Td]\) in Remark~\ref{rem:fut-dyn}.

\noindent The discounted conditional-expectation value is a martingale. By Proposition~\ref{pro2.1}, matching its diffusion coefficients and initial value gives the same discounted replicating wealth. The discounted payoff is a finite linear combination of exponentials of Gaussian integrals, multiplied by deterministic factors, and hence is square-integrable. Its conditional-expectation martingale is uniformly integrable, which establishes admissibility.
Finally, equation~\eqref{eq:phi0-collateral-choice} follows from
\(V=\phiv^0\bar B^d\).
\end{proof}

\brem\label{rem:hedging-special-cases-new}
The collateralisation process \(\beta\) and the collateral currency \(i\) enter the hedge ratios through the deterministic factors \(A^{\beta,i}\) contained in \(I^{(1)},I^{(2)},I^{(3)}\). Full collateralisation is obtained directly by setting $\beta\equiv1$.

When \(\gamma=0\), equation~\eqref{eq:phi-f-collateral-choice} gives \(\phif=0\). The absence of a position in \(\text{\euro STR}\) futures follows from the absence of foreign-rate risk in the price process. When \(\gamma\ne0\), the position \(\phif\) hedges the foreign floating-rate leg of the differential swap, which is not generated by the choice of collateral currency.

At \(t=\Ud\), \(\btU=\hbtU=0\), and the pre-accrual and accrual-period expressions agree. Hence both futures positions are continuous at the beginning of the accrual period.
\erem

\subsection{Hedging of multi-period differential swaps}
\label{sec:hedge-mp-collateral-choice}
We extend the preceding single-period hedge to the multi-period differential swap priced in Section~\ref{sec:multi-period-foreign-coll}. For each \(j=1,\ldots,n\), let \(\nu^{d,j}\) and \(\nu^{f,q,j}\) denote the diffusion coefficients of the SOFR and domestic-currency \(\text{\euro STR}\) futures gains processes associated with \([T_{j-1},T_j]\), so that
\[
\diff\Fd_t(T_{j-1},T_j)
=\nu^{d,j}_t\diff\Zone_t,
\qquad
\diff\Ffq_t(T_{j-1},T_j)
=\nu^{f,q,j}_t\diff\Ztwo_t,
\]
where \(\nu^{f,q,j}_t=Q_t\nu^{f,j}_t\).
The coupon-matched futures menu is an idealised assumption. Implementation requires a corresponding choice and rolling of the listed contracts.

\bp\label{pro:hedge-mp-collateral-choice}
The multi-period differential swap priced in Proposition~\ref{pro:mp-sofr-foreign-collat} can be replicated by trading, for each remaining coupon \(j=k(t),\ldots,n\), in the SOFR and \(\text{\euro STR}\) futures contracts associated with \([T_{j-1},T_j]\). Its net domestic cash position and collateral balance are
\begin{equation}\label{eq:phi0-multi-collateral-choice}
\phiv^0_t
=\frac{\Swapb_t(T_0,n;i)}{\bar B^d_t},
\qquad
C_t^i=-\beta_t\Swapb_t(T_0,n;i).
\end{equation}
The actual bank holdings follow from~\eqref{eq:actual-cash-holdings}, with $V_t=\Swapb_t(T_0,n;i)$.
For every \(j=k(t),\ldots,n\), the position in the corresponding SOFR futures contract is
\begin{equation}\label{eq:phi-d-multi-collateral-choice}
\phiv^{d,j}_t
=
\frac{
\mathbf{1}_{\{t\le T_{j-1}\}}\sigma_{T_{j-1}}(t)I^{(1)}_j(t)
-\sigma_{T_j}(t)\big(
\gamma I^{(2)}_j(t)
+(1-\gamma+\delta_j\kappa)I^{(3)}_j(t)
\big)}
{\nu^{d,j}_t}.
\end{equation}
In particular, for \(t\in[T_{j-1},T_j)\),
\begin{equation}\label{eq:phi-d-multi-collateral-choice-accrual}
\phiv^{d,j}_t
=
\frac{
\delta_j\big(
\gamma I^{(2)}_j(t)
+(1-\gamma+\delta_j\kappa)I^{(3)}_j(t)
\big)}
{1+\delta_j\Fd_t(T_{j-1},T_j)}.
\end{equation}
The position in the corresponding \(\text{\euro STR}\) futures contract is
\begin{align}
\phiv^{f,j}_t
&=
-\frac{
\gamma\big(
\mathbf{1}_{\{t\le T_{j-1}\}}\wh{\sigma}_{T_{j-1}}(t)
-\wh{\sigma}_{T_j}(t)
\big)I^{(2)}_j(t)}
{\nu^{f,q,j}_t} \nonumber\\
&=
-\frac{\gamma\delta_j}{Q_t}
A^{\beta,i}_{t,T_j}\Bd(t,T_j)
\Gamma^{\mathrm{qf}}_t(T_{j-1},T_j).
\label{eq:phi-f-multi-collateral-choice}
\end{align}
All futures positions corresponding to coupons already paid are set equal to zero.

Writing $F^{d,j}_t:=\Fd_t(T_{j-1},T_j)$ and $F^{f,j}_t:=\Ff_t(T_{j-1},T_j)$ for the corresponding futures, the replicating wealth between payment dates satisfies
\begin{equation}\label{eq:wealth-multi-collateral-choice}
\diff V_t=r^{\beta,i}_tV_t\diff t+\sum_{j=k(t)}^n\Big(\phiv^{d,j}_t\diff F^{d,j}_t+\phiv^{f,j}_tQ_t\diff F^{f,j}_t+\phiv^{f,j}_t\diff\qvar{Q,F^{f,j}}{t}\Big).
\end{equation}

At each payment date \(T_j\), the \(j\)th component of the portfolio delivers the corresponding coupon and is then closed. The ex-dividend wealth therefore satisfies $\Delta V_{T_j}=-\SwapT_{T_j}(T_{j-1},T_j)$, and $V_{T_n}=0$ after the last payment. In particular, if \(\gamma=0\), then \(\phiv^{f,j}=0\) for every remaining coupon.
\ep

\begin{proof}
For each \(j\), apply Proposition~\ref{pro:hedge-sp-sofr-partial-foreign} with
\((\Ud,\Td,\delta)=(T_{j-1},T_j,\delta_j)\). This gives equations~\eqref{eq:phi-d-multi-collateral-choice}--\eqref{eq:phi-f-multi-collateral-choice}. Since the pricing operator and the self-financing gains equation are linear, summing the single-period strategies over \(j=k(t),\ldots,n\) yields the value in~\eqref{eq:multi-price-new} and the wealth dynamics in~\eqref{eq:wealth-multi-collateral-choice}.
Moreover, the coupon-level balances add pointwise:
\[
\sum_{j=k(t)}^n C^j_t(i)
=-\beta_t\sum_{j=k(t)}^n V^j_t
=-\beta_t\Swapb_t(T_0,n;i).
\]

This gives~\eqref{eq:phi0-multi-collateral-choice}. At \(T_j\), the value of the \(j\)th subportfolio equals the \(j\)th contractual cash flow, so its delivery leaves the remaining subportfolios equal to the ex-dividend value of the outstanding coupons.
The discounted wealth including the payments already delivered is
\[
(B^\beta_t)^{-1}V_t
+\sum_{j:\,T_j\le t}(B^\beta_{T_j})^{-1}
\SwapT_{T_j}(T_{j-1},T_j).
\]
As a finite sum of square-integrable martingales, it is uniformly integrable. Hence the multi-period strategy is admissible in the sense stated in Section~\ref{def2.5}.
\end{proof}
\section{Numerical Studies}\label{sec:numerics}
In this section we use the constant collateralisation process $\beta_t\equiv\beta\in[0,1]$. The numerical study validates the closed-form pricing formulas, quantifies the collateral-carry and contractual quanto effects separately, and evaluates the futures hedges under discrete rebalancing. In particular, the numerical study separates the two channels identified in Sections~\ref{sec:pricing} and~\ref{sec:hedging}. The collateral currency determines the deterministic factors \(A^{\beta,i}_{t,T_j}\), whereas the foreign floating leg, when \(\gamma\ne0\), determines the stochastic foreign-rate exposure and the quanto correction. We therefore consider the four configurations obtained from \(\gamma\in\{0,1\}\) and \(i\in\{d,f\}\). The case \(\gamma=0\) is a domestic-payoff control, while \(\gamma=1\) is a SOFR--\(\text{\euro STR}\) differential swap. Unless otherwise stated, payments are semiannual and collateralisation is full, so that \(\beta=1\).

The specification is intended as a transparent Gaussian benchmark rather than a calibration to a complete market term structure. In particular, deterministic collateral spreads exclude stochastic cross-currency-basis and collateral-switching convexity, and the one-factor short-rate models exclude slope and twist risks. Within this scope, the closed-form expressions permit the effects of collateral carry and of the contractual foreign leg to be identified separately and provide an exact benchmark for numerical implementations of richer models.

\subsection{Numerical results}\label{subsec:numerics-validation-new}

The baseline parameters are reported in Table~\ref{tab:baseline-r4b}. All rate-dynamics parameters are reported under the domestic pricing measure \(\Q\). In particular, the foreign Vasicek drift constant is \(\wh c=0.057915\); equivalently, \(\wh a=\wh c+\wh\sigma\barsig\rhof=0.05799\) under \(\whQ\). For USD collateral we set the net carry spread \(\alpha^{c,d}-\alpha^d\) equal to zero. For EUR collateral we set \(\alpha^{c,f}-\alpha^f=20\) basis points. Consequently, the two collateral specifications differ only through their deterministic net carry.

\begin{table}[!htbp]
\centering
\caption{Baseline parameters for the numerical study.}
\label{tab:baseline-r4b}
\small
\begin{tabular}{lcc}
\toprule
Quantity & USD/SOFR & EUR/$\text{\euro STR}$ \\
\midrule
Initial overnight rate & $3.670\%$ & $1.933\%$ \\
Mean-reversion coefficient & $3.000$ & $3.000$ \\
Short-rate volatility & $1.200\%$ & $1.000\%$ \\
Vasicek drift constant& $0.1101$ & $0.057915$ \\
Net collateral carry $\alpha^{c,i}-\alpha^i$ & $0$ bp & $20$ bp \\
\midrule
\multicolumn{3}{l}{\footnotesize $\barsig=7.5\%$, $\rho_{12}=0.25$, $\rho_{13}=-0.25$, $\rho_{23}=0.10$, and $Q_0=1.1782$.} \\
\bottomrule
\end{tabular}
\end{table}

We validate Proposition~\ref{pro:single_sofr_foreign_collat} and Proposition~\ref{pro:mp-sofr-foreign-collat} using Monte Carlo draws from the exact joint Gaussian law of the integrated domestic and foreign rates. Thus the comparison contains sampling error but no Euler discretisation error. Table~\ref{tab:mc-validation-r4b} gives representative cases; all estimates are within two Monte Carlo standard errors of their closed-form values.

\begin{table}[!htbp]
\centering
\caption{Closed-form prices against Monte Carlo estimates.}
\label{tab:mc-validation-r4b}
\small
\begin{tabular}{lcccccc}
\toprule
Structure & $T$ & $\gamma$ & Collateral & $\beta$ & MC error & $z$-score \\
\midrule
Single & 0.5 & 0.0 & USD & 1.0 & $5.66e-06$ & $1.89$ \\
Single & 1.5 & 1.0 & EUR & 1.0 & $-6.78e-06$ & $-1.59$ \\
Multi & 3.0 & 0.0 & USD & 1.0 & $1.01e-05$ & $0.83$ \\
Multi & 3.0 & 1.0 & EUR & 1.0 & $-3.37e-06$ & $-0.25$ \\
\multicolumn{7}{l}{\footnotesize Exact-Gaussian Monte Carlo with 250,000 paths; MC error is MC minus closed form.} \\
\bottomrule
\end{tabular}

\end{table}

Additional numerical identities provide checks on the structure of the formulas. The single-period par margin is invariant to both \(i\) and \(\beta\). When \(\gamma=0\), changing all foreign-rate and FX parameters leaves the price unchanged, provided the collateral net-carry specifications are held fixed. Finally, two collateral currencies with the same net carry \(\alpha^{c,i}-\alpha^i\) give identical values. These identities test the separation between collateral carry and contractual cross-currency risk independently of the Monte Carlo comparison.

\subsection{Collateral currency as a deterministic carry adjustment}\label{subsec:numerics-collateral-carry}

To compare collateral currencies, we value the same contract under the two collateral specifications. For a fixed margin \(\kappa\), define the collateral-currency value adjustment
\begin{equation}\label{eq:numerical-csa-switch}
\Delta V^{f-d}_0(\kappa)
:=\Swapb_0(T_0,n;f)-\Swapb_0(T_0,n;d).
\end{equation}
This differs from comparing two par contracts, since the latter generally have different fixed margins. Equation~\eqref{eq:multi-price-new} shows that \(\Delta V^{f-d}_0\) is generated by the alternative deterministic weights \(A^{\beta,f}_{0,T_j}\) and \(A^{\beta,d}_{0,T_j}\).

For completeness, Table~\ref{tab:collateral-summary-r4b} reports both the change in the par margin and the value adjustment for a contract whose fixed margin is 100 basis points above the USD-collateral par margin. The par-margin changes are very small in the present baseline: for a contract struck at par, the expected coupon differences are already close to balanced at each payment date, so changing their deterministic weights produces substantial cancellation. This observation is a feature of the cash-flow profile and is not a general invariance result for multi-period contracts.

\begin{table}[!htbp]
\centering
\caption{Effect of replacing USD collateral by EUR collateral.}
\label{tab:collateral-summary-r4b}
\small
\begin{tabular}{ccrr}
\toprule
$\gamma$ & Maturity & Par-margin change (bp) & CSA-switch value (USD) \\
\midrule
0 & 5Y & 0.000035 & 24,129 \\
0 & 10Y & 0.000041 & 81,503 \\
1 & 5Y & -0.000042 & 24,125 \\
1 & 10Y & -0.000049 & 81,495 \\
\midrule
\multicolumn{4}{l}{\footnotesize EUR minus USD collateral, full collateralisation. The CSA-switch value uses} \\
\multicolumn{4}{l}{\footnotesize USD 100 million notional and a fixed margin 100 bp above the USD-collateral par margin.} \\
\bottomrule
\end{tabular}

\end{table}

\begin{figure}[!htbp]
\centering
\includegraphics[width=0.82\textwidth]{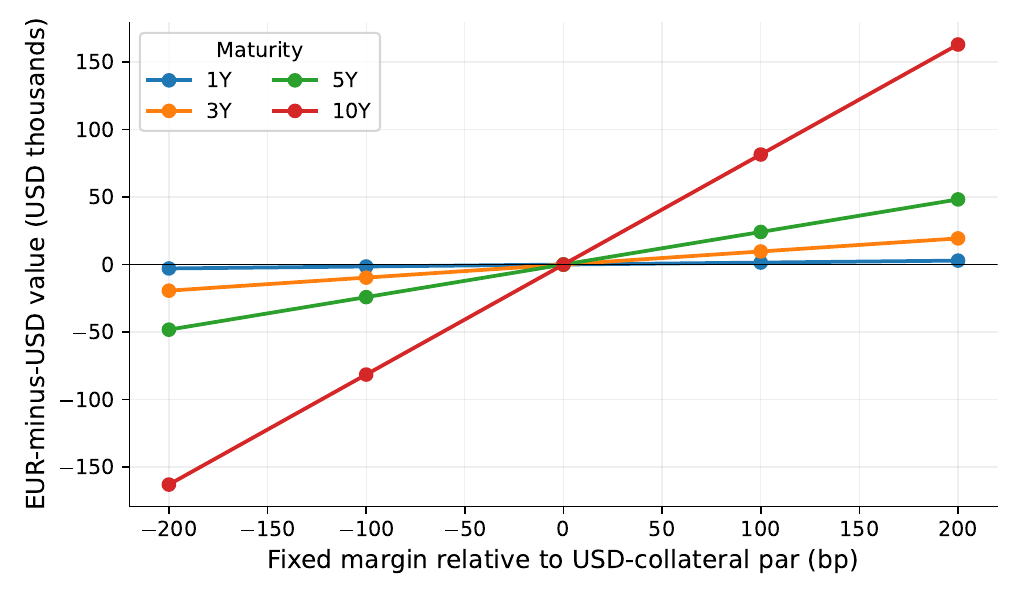}
\captionsetup{skip=12pt}
\caption{EUR-collateral value minus USD-collateral value for the same fully collateralised differential swap, per USD 100 million notional.}
\label{fig:collateral-value-transfer}
\end{figure}

Figure~\ref{fig:collateral-value-transfer} considers a genuine differential swap with \(\gamma=1\) and shows \(\Delta V^{f-d}_0\) for fixed margins between 200 basis points below and 200 basis points above the USD-collateral par margin. Values are stated per USD 100 million notional. The adjustment is close to zero for an at-par contract but becomes economically visible for an existing off-market position. At 100 basis points above par, it is approximately USD \(24{,}125\) for a five-year contract and USD \(81{,}495\) for a ten-year contract. Since EUR collateral has the higher net carry in this experiment, it has a larger discounting effect; the sign of the adjustment therefore reverses with the sign of the off-market position.

The adjustment in Figure~\ref{fig:collateral-value-transfer} is a discounting effect. It is not driven by FX volatility or by an additional foreign-rate state variable in the collateral account. Under the present fixed-currency convention, collateral currency selection at inception is therefore a static comparison of deterministic net-carry curves. Dynamic substitution among eligible currencies would constitute a distinct cheapest-to-deliver problem and lies outside the present framework.

\subsection{Contractual quanto adjustment}\label{subsec:numerics-quanto}

We next isolate the stochastic cross-currency effect generated by the contractual foreign leg. For collateral currency \(i\) and contractual foreign-leg weight \(\gamma\), define
\begin{equation}\label{eq:numerical-quanto-adjustment}
\Delta\kappa^{\mathrm{q}}_i(\gamma)
:=\kappa^\star_0(T_0,n;\beta,\gamma,i)
-\kappa^{\star,\mathrm{no\text{-}qf}}_0(T_0,n;\beta,\gamma,i),
\end{equation}
Here \(\kappa^{\star,\mathrm{no\text{-}qf}}_0\) is the counterfactual par margin obtained from~\eqref{eq:par-swap-rate-mp-diff} by replacing the multiplicative correction \(\Gamma^{\mathrm{qf}}\) by its neutral value \(1\), or equivalently by setting the exponent in~\eqref{eq:quanto-correction} equal to zero. The actual par margin \(\kappa^\star_0\) uses the calculated value of \(\Gamma^{\mathrm{qf}}\). Hence
\begin{align}
\Delta\kappa^{\mathrm{q}}_i(\gamma)
&=-\gamma
\frac{\displaystyle\sum_{j=1}^n
A^{\beta,i}_{0,T_j}\Bd(0,T_j)
\big(1+\delta_j\Ff_0(T_{j-1},T_j)\big)
\big(\Gamma^{\mathrm{qf}}_0(T_{j-1},T_j)-1\big)}
{\displaystyle\sum_{j=1}^n\delta_j I^{(3)}_j(0)}
\nonumber\\
&=\gamma\Delta\kappa^{\mathrm{q}}_i(1).
\label{eq:quanto-adjustment-gamma-scaling}
\end{align}
The quanto adjustment for any contractual value of \(\gamma\) follows exactly by scaling. This construction removes only the covariance correction and retains the traded \(\text{\euro STR}\)-futures rate.

\begin{figure}[!htbp]
\centering
\includegraphics[width=0.98\textwidth]{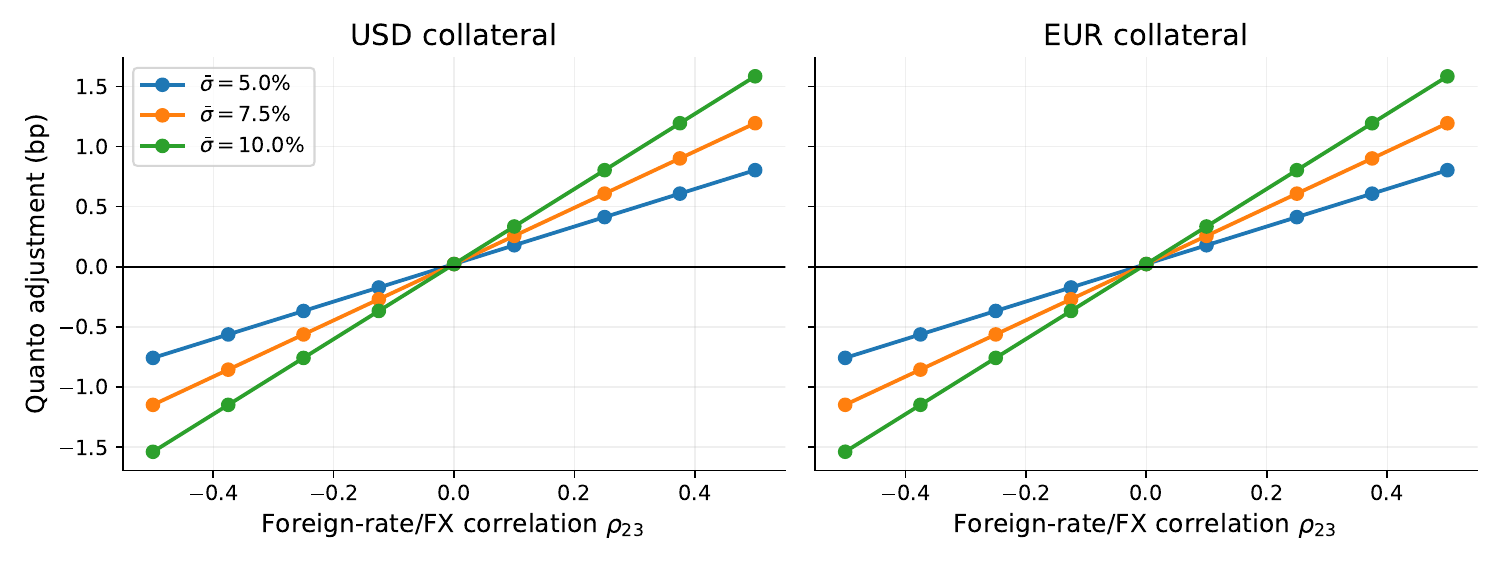}
\caption{Quanto adjustment to the par margin of a five-year differential swap with \(\gamma=1\), under USD collateral (left) and EUR collateral (right).}
\label{fig:quanto-adjustment-r4b}
\end{figure}

Figure~\ref{fig:quanto-adjustment-r4b} reports \(\Delta\kappa^{\mathrm{q}}_i(1)\) for a five-year differential swap as a function of the foreign-rate/FX correlation \(\rho_{23}\) and the FX volatility \(\bar\sigma\), under both USD and EUR collateral. At the baseline values \(\rho_{23}=0.10\) and \(\bar\sigma=7.5\%\), the quanto adjustments are \(0.2570\) and \(0.2569\) basis points under USD and EUR collateral, respectively. Holding the FX volatility at \(7.5\%\), the USD-collateral adjustment ranges from approximately \(-1.15\) to \(1.19\) basis points as \(\rho_{23}\) varies from \(-0.5\) to \(0.5\); the corresponding EUR-collateral range is virtually indistinguishable.
In this sensitivity experiment, the foreign-measure drift constant $\wh a$ is held fixed. Thus the foreign futures curve is unchanged, while the domestic-measure drift constant $\wh c=\wh a-\wh\sigma\barsig\rhof$ varies with $\barsig$ and $\rhof$.

The two panels exhibit the same stochastic risk channel because the quanto correction is generated by the contractual foreign overnight-rate accrual. Changing the collateral currency modifies only the deterministic payment-date weights and therefore produces the small difference between the panels. The correction is absent when \(\gamma=0\), irrespective of the collateral currency. Moreover, \(Q_0\) and \(\rho_{13}\) do not enter the price of the present fixed-conversion differential swap, since its contractual cash flows contain no random FX amount. Hence the sensitivity in Figure~\ref{fig:quanto-adjustment-r4b} is attributable to the contractual quanto component rather than to foreign collateralisation.

\subsection{Discrete implementation of the futures hedges}\label{subsec:numerics-hedging-new}\label{subsec:numerics-hedging}

We finally implement the strategies of Proposition~\ref{pro:hedge-mp-collateral-choice} by integrating the reduced wealth equation~\eqref{eq:wealth-multi-collateral-choice} on discrete rebalancing grids. The experiment uses a three-year semiannual swap, 10,000 common paths simulated under \(\Q\), and daily Euler--Maruyama state evolution with \(\Delta t=1/252\) and left-end sums for the rate integrals. The foreign-futures gain over each daily interval is converted using the right-end rule \(Q_{t_{k+1}}\Delta\Ff_{t_{k+1}}\), consistently with Section~\ref{sec:collat-futures}. At each payment date the corresponding coupon is delivered and its futures positions are closed. The terminal post-payment wealth is therefore the replication error.

We compare an unhedged position, a SOFR-futures hedge, and the full SOFR--\(\text{\euro STR}\) futures hedge. The SOFR-only strategy uses the domestic position of the full hedge and sets its foreign position to zero. The fixed margin is the USD-collateral par margin and is held fixed when the collateral currency is changed. Table~\ref{tab:hedging-r4b} reports daily-rebalancing results. The variance ratio is computed relative to the unhedged terminal error.

\begin{table}[!htbp]
\centering
\caption{Terminal hedging errors under daily rebalancing.}
\label{tab:hedging-r4b}
\small
\begin{tabular}{cclrr}
\toprule
$\gamma$ & Collateral & Hedge & RMSE (bp) & Variance ratio \\
\midrule
0 & USD & SOFR only & 0.020 & $9.04\times 10^{-8}$ \\
0 & USD & SOFR+$\text{\euro STR}$ & 0.020 & $9.04\times 10^{-8}$ \\
0 & EUR & SOFR only & 0.020 & $9.28\times 10^{-8}$ \\
0 & EUR & SOFR+$\text{\euro STR}$ & 0.020 & $9.28\times 10^{-8}$ \\
1 & USD & SOFR only & 55.399 & 0.526390 \\
1 & USD & SOFR+$\text{\euro STR}$ & 0.263 & $1.19\times 10^{-5}$ \\
1 & EUR & SOFR only & 55.533 & 0.526391 \\
1 & EUR & SOFR+$\text{\euro STR}$ & 0.264 & $1.19\times 10^{-5}$ \\
\multicolumn{5}{l}{\footnotesize Three-year semiannual swap, daily rebalancing, 10,000 common simulated paths.} \\
\bottomrule
\end{tabular}

\end{table}

For \(\gamma=0\), the \(\text{\euro STR}\)-futures position is identically zero under both collateral currencies. Consequently, the SOFR-only and full strategies coincide; their daily RMSE is approximately \(0.020\) basis points of notional. For \(\gamma=1\), the SOFR-only strategy leaves the foreign floating-rate risk unhedged and has an RMSE of about \(55.4\) basis points. Adding the \(\text{\euro STR}\) futures reduces the RMSE to approximately \(0.263\) basis points and the residual variance to about \(1.2\times10^{-5}\) of the unhedged variance. The corresponding results under USD and EUR collateral are nearly identical in relative terms: changing the collateral currency rescales the hedge ratios through \(A^{\beta,i}\), but it does not change the set of stochastic factors that must be hedged.

\begin{figure}[!htbp]
\centering
\includegraphics[width=0.82\textwidth]{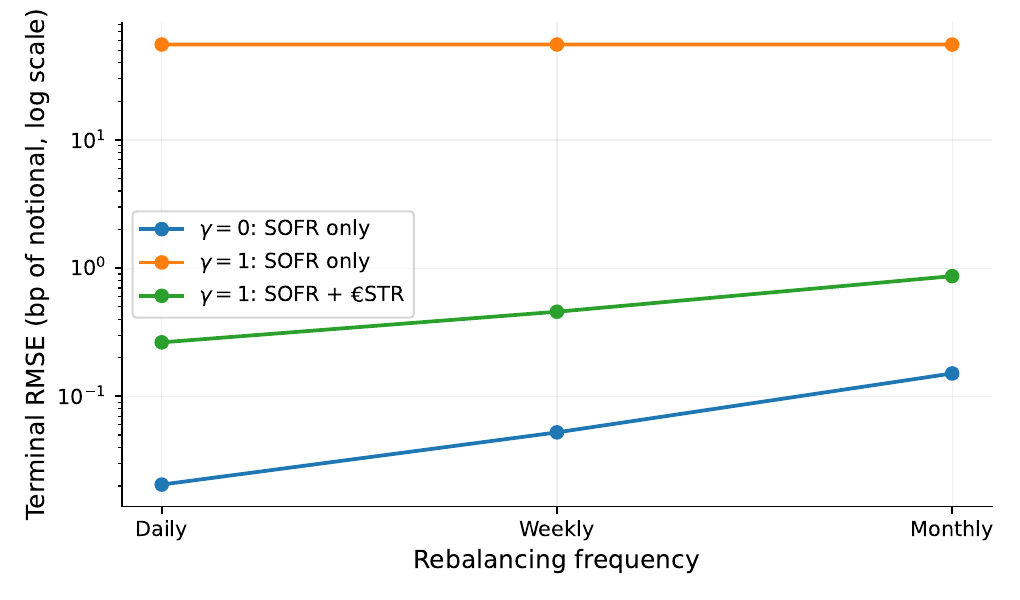}
\caption{Terminal hedging RMSE under EUR collateral for daily, weekly, and monthly rebalancing. The vertical scale is logarithmic.}
\label{fig:hedging-frequency-r4b}
\end{figure}

Figure~\ref{fig:hedging-frequency-r4b} shows the effect of reducing the rebalancing frequency under EUR collateral. For \(\gamma=0\), the SOFR hedge remains the full model hedge. For \(\gamma=1\), the SOFR-only error changes little with rebalancing frequency because it is dominated by the omitted foreign factor. The two-futures RMSE increases from \(0.264\) basis points under daily rebalancing to \(0.866\) basis points under monthly rebalancing, as expected from discretising a continuous-time replicating strategy.

Taken together, the numerical results confirm the separation established analytically. Collateral currency affects valuation and hedge magnitudes through deterministic net carry and payment-date weights. The foreign floating leg determines the quanto adjustment and the need for \(\text{\euro STR}\) futures. In particular, selecting EUR rather than USD collateral does not by itself create a foreign-rate hedge requirement under the retained-principal convention.
\section{Conclusions}\label{sec:conclusion}

This paper has developed a valuation and futures-based hedging framework for USD-settled differential swaps under proportional collateralisation in either USD or EUR. The hedger has direct access to borrowing and investment in both currencies at symmetric money-market rates. If collateral is posted or received in EUR, interest cash flows are settled against the USD account through spot conversion. For USD collateral, the principal is already included in the USD account. In either case, the account equations yield a domestic net-wealth equation driven by futures gains and collateral net carry. For collateral currency \(i\), this gives the effective rate
\[
r^{\beta,i}_t=\bar r^d_t+\beta_t\big(r^{c,i}_t-\bar r^i_t\big).
\]
The collateral benchmark rate cancels between the return on the retained principal and the collateral remuneration, so deterministic spreads produce a deterministic adjustment to domestic discounting.

Under a tractable Gaussian cross-currency specification, we obtained closed-form prices for single- and multi-period SOFR/\(\text{\euro STR}\) differential swaps. The formulas separate the collateral-currency channel from the contractual cross-currency channel. The selected collateral currency determines the net-carry factors that weight cash flows at their payment dates. The foreign floating-rate leg instead generates the quanto correction. This distinction is invisible for the par margin of a single-period swap, where the common collateral factor cancels, but it can affect a multi-period value because cash flows paid at different dates receive different weights.

The same separation determines the hedge. Collateral currency changes the magnitude of the hedge ratios but, under deterministic spreads, does not introduce an additional Brownian factor. When \(\gamma=0\), the payoff has no foreign-rate exposure and the \(\text{\euro STR}\)-futures position is zero under both USD and EUR collateral. When \(\gamma\ne0\), the foreign-futures position hedges the contractual \(\text{\euro STR}\) leg. The numerical experiments validate the closed-form prices against Monte Carlo estimates and confirm the replication results under discrete rebalancing. They also show that a collateral-currency change may have only a small effect on the par margin in the baseline configuration while producing an economically visible value transfer for an existing off-market position.

The model is intended as an exact analytical benchmark rather than a complete market calibration. The unshifted Vasicek models cannot fit arbitrary observed initial term structures, although deterministic shifts may be introduced to fit the initial curves without altering the one-factor diffusion structure. Its one-factor rate dynamics omit slope and twist risks, while deterministic funding and collateral spreads exclude stochastic basis risk and collateral-switching convexity. These restrictions make the two economic channels transparent and provide controlled benchmarks for richer implementations. Natural extensions include multi-factor Gaussian or affine term-structure models, stochastic collateral and funding spreads, and a dynamic substitution rule for several eligible collateral currencies. Transaction costs, asymmetric borrowing and lending rates, thresholds, and counterparty default would lead beyond the present linear framework and are also relevant directions for further work.

\vskip 10 pt
\noindent{\bf Acknowledgements.}
The authors thank the Editor and the anonymous referees for their constructive comments, which helped improve the manuscript and highlighted useful directions for future research. The research of R. Liu and M. Rutkowski was supported by the Australian Research Council Discovery Project scheme under grant DP200101550 {\it Fair pricing of superannuation guaranteed benefits under downturn risk}. This work was also supported in part by the National Natural Science Foundation of China under Grant 12371447.

\noindent{\bf Disclosure of interest.}
No potential conflict of interest was reported by the author(s).

\noindent\textbf{Declaration of Generative AI Use.} The authors used ChatGPT (OpenAI; GPT-6.1 Sol) to assist with language editing and consistency checks of mathematical derivations, and development and checking of numerical code. These tools were used to improve clarity and support the revision and numerical implementation. The authors reviewed and revised the AI-assisted material, confirmed the originality and accuracy of the final manuscript, and take full responsibility for its content. The authors have checked the applicable terms of use and confirmed the suitability of the assisted material for publication.

\bibliographystyle{plain}
\bibliography{currencybiblio}

\end{document}